\documentclass[prb,preprint,showpacs]{revtex4-1}

\usepackage[dvipdfmx]{graphicx}

\usepackage{amsmath, amssymb}
\usepackage[cdot,mediumqspace,Grey,squaren]{SIunits}
\usepackage{color}
\usepackage{caption}
\usepackage[ps2pdf]{hyperref}

\renewcommand{\vec}{\mathbf} 
\newcommand{\Ham}{\hat{H}} 
\newcommand{\e}[1]{\text{e}^{#1}} 
\renewcommand{\k}{\vec{k}} 
\newcommand{\diffsec}[3][2]{\dfrac{\partial^#1 #2}{\partial #3^#1 }} 
\newcommand{\Dr}[2][3]{d^{#1} #2} 
\newcommand{\db}[1]{\overline{\overline{#1}}} 
\renewcommand{\Re}{\text{Re}} 
\renewcommand{\Im}{\text{Im}} 

\graphicspath{{./Figures_for_submission/},{./}}

\begin{document}

\title{Effective Hamiltonian for electron waves in artificial graphene: A first principles derivation}

\author{Sylvain Lanneb\`{e}re}
\author{M\'{a}rio G. Silveirinha}
\affiliation{University of Coimbra, Department of Electrical Engineering -- Instituto de Telecomunica\c{c}\~{o}es, Coimbra 3030-290, Portugal}

\begin{abstract}
We propose a first principles effective medium formalism to study
the propagation of electron waves in semiconductor heterostructures
with a zero-band gap. Our theory confirms that near the $K$-point
the dynamics of a two-dimensional electron gas modulated by an
external electrostatic potential with honeycomb symmetry is
described by the same pseudospinor formalism and Dirac massless
equation as a graphene monolayer. Furthermore, we highlight that
even though other superlattices based on semiconductors with a
zincblende-type structure can have a zero band-gap and a linear
energy-momentum dispersion, the corresponding effective medium
Hamiltonian is rather different from that of graphene, and can be
based on a single-component wavefunction.
\end{abstract}


\date{\today}

\maketitle

\section{Introduction}

The experimental discovery of graphene in 2004
\cite{novoselov_electric_2004} opened the door to a large scientific
activity, and nowadays graphene physics is one of the most vibrant
research fields in condensed matter physics. One of the fascinating
electronic properties of graphene is that low energy electrons may
be described by a 2D massless Dirac equation. Consequently, the
electron states have a linear energy dispersion with a zero band gap
\cite{castro_neto_electronic_2009,das_sarma_electronic_2011}. This
feature is interesting not only because it may enable ultrafast
carbon-based electronics due to the high electron mobility,
but also to create tunable ``one-atom-thick'' platforms for infrared photonic functions \cite{engheta_science, abajo_1, abajo_2, novoselov_plasmonics}
with a strong non linear optical response \cite{mikhailov_nonlinear_2008}.

Recently, other mechanisms have been suggested to mimic the
extraordinary properties of graphene, such as using ultracold atoms
in hexagonal or honeycomb optical lattices
\cite{zhu_simulation_2007,wu_flat_2007,wu_pxy-orbital_2008,wunsch_dirac-point_2008},
using photonic crystals (``photonic graphene'')
\cite{segev_graphene}, or applying a periodic potential onto a
two-dimensional electron gas (2DEG). This latter idea was first
described in Ref. \cite{park_making_2009}, where the authors studied
a nanopatterned electron gas with hexagonal symmetry, and
demonstrated that the electrons behavior is governed by a massless
Dirac equation. The case of a potential with honeycomb symmetry was
considered in Ref. \cite{gibertini_engineering_2009}, which also
reported a possible realization based on modulation-doped GaAs
quantum wells.

The conclusion that electrons may behave as massless Dirac fermions
in a 2DEG modulated by an external potential is largely rooted in
the observation that the energy dispersion is linear near the
corners of the Brillouin zone and on an analogy with graphene
\cite{park_making_2009, gibertini_engineering_2009}. In our view, it
would be desirable to have a more solid theoretical foundation of
such an important result, and have a more complete understanding if
linear dispersing bands do always imply that the time dynamics of
the electron waves is described by a massless Dirac equation.
Indeed, one may wonder if a different type of physics may be as well
compatible with linear dispersing bands.

With this objective in mind, in this article we develop a first
principles approach to characterize the dynamics of electron waves
in artificial graphene nanomaterials. Our theory is based on the
effective medium theory for electron waves developed in Ref.
\cite{silveirinha_effective_2012}. In that work, it was shown that
it is possible to introduce an effective Hamiltonian that describes
exactly the time evolution of electron states that vary slowly on
the scale of the unit cell. Furthermore, the electronic band
structure obtained from the effective medium Hamiltonian is exactly
coincident with what is found from the microscopic Hamiltonian
\cite{silveirinha_effective_2012}. Here, we apply an extended
version of this theory to demonstrate from first principles that the
electron wavefunction envelope in a nanopatterned 2DEG with
honeycomb symmetry satisfies, indeed, the 2D massless Dirac
equation. Moreover, we also investigate the effective medium
description of superlattices\cite{Esaki} based on semiconductors
with zincblende-type structure. Consistent with our previous studies
\cite{silveirinha_transformation_electronics,
silveirinha_metamaterial-inspired_2012, silveirinha_giant_2014}, it
is found the electron energy dispersion may be linear for some
structural parameters. Interestingly, it is shown that in this
system the electrons do not have a pseudospin degree of freedom,
quite different from what happens in graphene wherein the electrons
are chiral fermions.

This paper is organized as follows. In section
\ref{sec:honeycomb_lattice}, we present a brief overview of the
theory of Ref. \cite{silveirinha_effective_2012}, and apply it to
characterize a 2DEG modulated by an external electrostatic potential
with honeycomb symmetry. It is shown that a direct application of
the method yields a single component Hamiltonian characterized by a
strongly  nonlocal (spatially dispersive) response. To circumvent
this problem, in section \ref{sec:pseudospinor} the effective medium
approach is extended to allow for a pseudospinor formalism. It is
proven that, similar to graphene, the electrons in the superlattice
behave as Dirac fermions and are described by the massless Dirac
equation. A parametric study of the influence of the external
potential amplitude and of the geometry on the effective Hamiltonian
is presented. In section \ref{sec:triangular_superlattice}, we
analyze an hexagonal superlattice formed by
mercury-cadmium-telluride, and prove that the time evolution of
electron waves in this second platform can be done based on a single
component wavefunction. Finally, in section \ref{sec:conclusion} the
conclusions are drawn.

\section{Single-component Hamiltonian for a potential with the honeycomb symmetry} \label{sec:honeycomb_lattice}

In this section, we describe how to obtain the single component
effective medium Hamiltonian for an artificial graphene platform
based on a two-dimensional electron gas, and discuss the
characteristics and limitations of such a description.

\subsection{Overview of the effective medium approach}
\label{sec:homogenisation_scheme}

In the following, we present a brief overview of the effective
medium approach developed in Ref. \cite{silveirinha_effective_2012}. The
use of effective medium concepts (e.g. the effective mass) has a
long tradition in condensed matter physics, and some relevant works
can be found in Refs. \cite{bastard_wave_1988, voon_k_2009, burt_1,
burt_2, foreman_1, pereira_mauro}.

To begin with, we consider a periodic system (e.g. a semiconductor
superlattice) described at the microscopic level by the Hamiltonian
$\Ham$ and whose time evolution is determined by the Schr\"{o}dinger
equation
\begin{equation}
\Ham \psi(\vec{r},t) = i \hbar \frac{\partial }{\partial t}\psi(\vec{r},t).
\label{E:Schrodinger_equation}
\end{equation}
The key idea of the method is to introduce an effective Hamiltonian
$\Ham_\text{ef}$ that describes exactly the time evolution of
initial ($t=0$) ``macroscopic'' states,  through a homogenized
Schr\"{o}dinger equation
\begin{equation}
\left( \Ham_\text{ef} \Psi \right)(\vec{r},t)  = i \hbar \frac{\partial }{\partial t}\Psi(\vec{r},t).
\label{E:effective_Schrodinger_equation}
\end{equation}
We say that an electron state $\psi$ is macroscopic if it is
invariant after spatial averaging: $\psi(\vec{r}) = \left\{
\psi(\vec{r}) \right\}_\text{av}$. Here, we consider that the
spatial averaging operator $\left\{ ~ \right\}_\text{av}$ is
equivalent to an ideal band-pass spatial filter such that for a
generic function $g$ depending on the spatial variable $\vec{r}$
\begin{equation}
\left\{ g(\vec{r}) \right\}_\text{av} = \int_{-\infty}^\infty  g(\vec{r}-\vec{r}') w(\vec{r}') \Dr[N]{\vec{r}'},
\end{equation}
where $w$ is a test function whose Fourier transform $\tilde{w}(\k)=\int w(\vec{r}) \e{-i \k \cdot \vec{r}} \Dr[N]{\vec{r}}$ has the
following properties \cite{silveirinha_effective_2012, silveirinha_poynting_2009}
\begin{equation}
\tilde{w}(\k)=\left\{
\begin{tabular}{c}
 1, $\k \in$ BZ\\
 0, $\k \notin$ BZ
\end{tabular}\right. , \label{E:BZ}
\end{equation}
where $\k$ is the wavevector and BZ stands for the relevant
Brillouin zone. For example, if the states that determine the
physics of the system are near the $\Gamma$-point then BZ should be
taken as the first Brillouin zone. In the above, $N$ represents the
number of relevant spatial dimensions (in this article $N=2$). From
a physical point of view, the property $\psi(\vec{r}) = \left\{
\psi(\vec{r}) \right\}_\text{av}$ is equivalent to say that the
electron state cannot be more localized than the characteristic
spatial period of the system. The effective Hamiltonian
$\Ham_\text{ef}$ is defined so that if $\psi_{t=0}(\vec{r}) =
\left\{ \psi_{t=0}(\vec{r}) \right\}_\text{av}$, i.e. if the initial
time state is macroscopic, then the solutions of Eqs.
\eqref{E:Schrodinger_equation} and
\eqref{E:effective_Schrodinger_equation} with the same initial time
conditions are linked by $\Psi(\vec{r},t) = \left\{ \psi(\vec{r},t)
\right\}_\text{av}$. In other words, the effective Hamiltonian
describes the dynamics of the smooth part of the wave function
$\psi(\vec{r},t) $. A consequence of this
property is that for an initial macroscopic state:
\begin{equation}
\left( \Ham_\text{ef} \Psi \right)(\vec{r},t) = \left\{ \left( \Ham
\psi \right)(\vec{r},t) \right\}_\text{av}, \,\, t>0
\label{E:definition_effective_hamiltonian}
\end{equation}
where $\Psi(\vec{r},t) = \left\{ \psi(\vec{r},t)
\right\}_\text{av}$. Calculating the unilateral Laplace transform of
both sides of the equation (e.g. the Laplace transform of $\psi
(\vec{r}, t)$ is $\psi (\vec{r},\omega)= \int_0^\infty dt \,
\psi(\vec{r},t)\, \e{i\omega t } $) we get:
\begin{equation}
\left( \Ham_\text{ef} \Psi \right)(\vec{r},\omega) = \left\{ \left(
\Ham \psi \right)(\vec{r},\omega) \right\}_\text{av}
\label{E:definition_effective_hamiltonian_Fourier}.
\end{equation}
The above identity may be used to numerically determine the
effective Hamiltonian as detailed below.

It was shown in Ref. \cite{silveirinha_effective_2012} that the
action of the operator $H_\text{ef}$ on the macroscopic wave
function is given in the space and time domains by the convolution
\begin{equation}
\left(\Ham_\text{ef} \Psi\right)(\vec{r},t)  = \int \Dr[N]{\vec{r}'}
\int_0^t dt \, H_\text{ef}(\vec{r}-\vec{r}',t-t') \Psi(\vec{r}',t').
\label{E:convolution}
\end{equation}
The Fourier transform of the kernel $H_\text{ef}(\vec{r},t)$ is
denoted by $H_\text{ef}(\k,\omega)= \int \Dr[N]{\vec{r}}
\int_0^\infty dt \, H_\text{ef}(\vec{r},t)\, \e{i\omega t } \e{-i\k
\cdot \vec{r}}$. Clearly, the function $H_\text{ef}(\k,\omega)$
completely determines the effective Hamiltonian. To obtain an
explicit formula for $H_\text{ef}(\k,\omega)$, we calculate the
unilateral Laplace transform of the Schr\"{o}dinger equation
\eqref{E:Schrodinger_equation} to find that:
\begin{equation}
(\Ham-E) \psi(\vec{r},\omega) = -i \hbar\psi_{t=0}(\vec{r}),
\label{E:schrodinger_laplace}
\end{equation}
with $E=\hbar \omega$. Thus, applying the spatial averaging operator
to both sides of the equation and using
\eqref{E:definition_effective_hamiltonian_Fourier}, it follows that
for an initial macroscopic state:
\begin{equation}
(H_\text{ef} \Psi) (\vec{r},\omega) - E \Psi(\vec{r},\omega) = -i
\hbar\psi_{t=0}(\vec{r}). \label{E:averaged_Laplace_equation}
\end{equation}
Let us now consider the particular case wherein the initial state is
$\psi_{t=0}(\vec{r}) \sim \e{i\k\cdot \vec{r}}$. Clearly, in these
conditions the solution of Eq. \eqref{E:schrodinger_laplace} has the
$\k$-Bloch property. Functions with the Bloch property, with $\k$ in
the BZ, have spatial averages of the form
\cite{silveirinha_effective_2012}
\begin{equation}
\Psi(\vec{r}')= \psi_\text{av} \e{i\k \cdot \vec{r}'}
\label{E:average_psi_1}
\end{equation}
where
\begin{equation}
\psi_\text{av} = \frac{1}{V_c}\int_\Omega \psi(\vec{r}) \e{-i\k\cdot
\vec{r}} \Dr[N]{\vec{r}}, \label{E:average_psi_time}
\end{equation}
where ${V_c}$ is the volume of the unit cell $\Omega$ in the spatial
domain. This property and Eq. \eqref{E:convolution} imply that
$\left(\Ham_\text{ef} \Psi\right)(\vec{r},\omega) =
H_\text{ef}(\k,\omega) \Psi(\vec{r},\omega)$ and hence from Eq.
\eqref{E:definition_effective_hamiltonian_Fourier} we get:
\begin{equation}
H_\text{ef}(\k,\omega) \Psi(\vec{r},\omega) =  \left\{ \left( \Ham
\psi \right)(\vec{r},\omega) \right\}_\text{av}.
\label{E:link_mac_mic}
\end{equation}
Hence, by numerically solving Eq. \eqref{E:schrodinger_laplace} with
respect to $\psi$ for the initial macroscopic state
$\psi_{t=0}(\vec{r}) \sim \e{i\k\cdot \vec{r}}$ and feeding the
result to Eq. \eqref{E:link_mac_mic}, it is possible to compute the
unknown function $H_\text{ef}(\k,\omega)$ for any value of
$(\k,\omega)$. Note that $\left( \Ham \psi \right)(\vec{r},\omega)$
has the $\k$-Bloch property, and hence $\left\{ \left( \Ham \psi
\right)(\vec{r},\omega) \right\}_\text{av}$ can be obtained using a
formula analogous to Eq. \eqref{E:average_psi_1}.

One interesting property of the effective Hamiltonian is that the
solutions of
\begin{equation}
\det(H_\text{ef}(\k,\omega)-E)=0, \label{E:energy_states}
\end{equation}
yield the exact energy band structure of the original microscopic
Hamiltonian. For more details the reader is referred to Ref.
\cite{silveirinha_effective_2012}. In previous works, this general
formalism was applied to graphene and semiconductor superlattices in
different contexts \cite{silveirinha_transformation_electronics,
silveirinha_metamaterial-inspired_2012, silveirinha_giant_2014,
silveirinha_electron_lens, fernandes_wormhole}. Moreover, related
effective medium techniques have been widely used to model the
propagation of light in electromagnetic metamaterials
\cite{silveirinha_homogenization, silveirinha_homogenization_2,
alu_homogenization, shvets_homogenization}.

\subsection{Single-component Hamiltonian}
\label{sec:scalar_Hamiltonian}

\begin{figure}[!h]
    \centering
 \includegraphics[width=\linewidth]{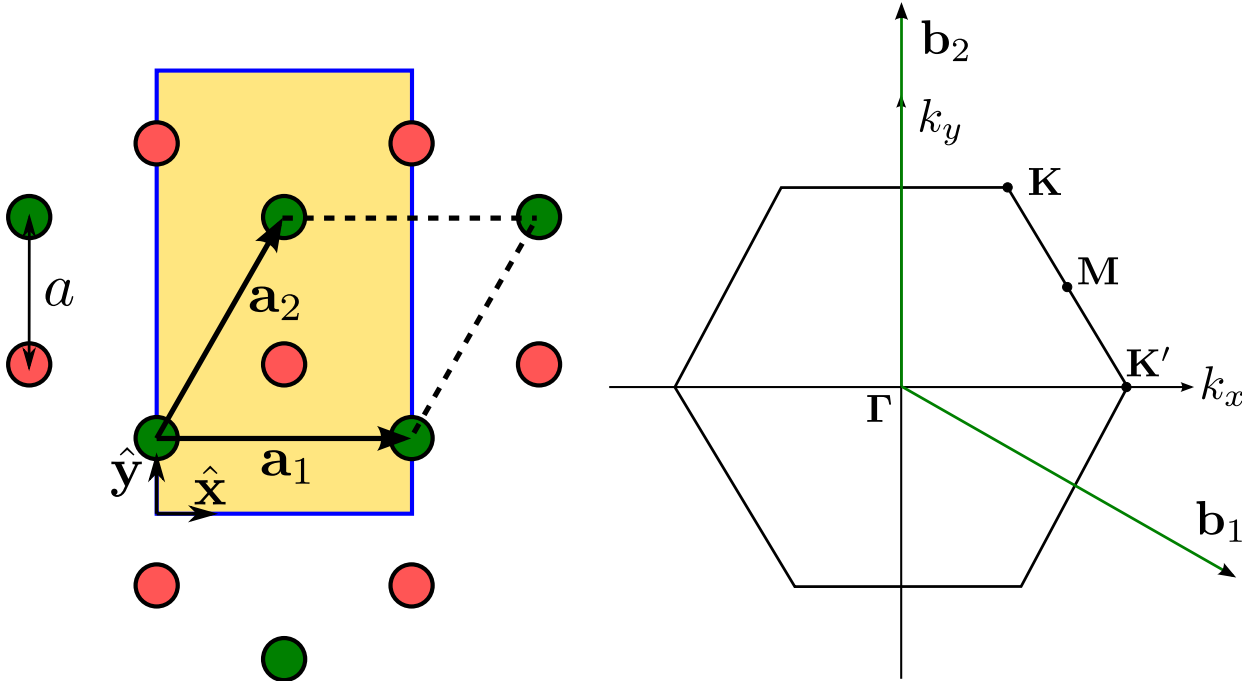}
          \caption{Left: 2DEG modulated by an applied potential with the honeycomb symmetry. A primitive cell is delimited by the dotted lines.
    The supercell used in the FDFD numerical calculations is represented by the coloured area. Right: First Brillouin zone with the usual high-symmetry
     points.}
\label{fig:honeycomb_lattice}
\end{figure}

Next, we apply the formalism described in the previous subsection to
characterize a 2DEG under the action of a periodic electrostatic
potential $V(\vec{r})$ with the honeycomb symmetry. Similar to Ref.
\cite{gibertini_engineering_2009}, it is assumed that the system
corresponds to a modulation-doped GaAs/AlGaAs quantum well. The
geometry of the patterned 2DEG is represented in Fig.
\ref{fig:honeycomb_lattice} (left). It is assumed that the electric
potential is a constant $V_0$ inside each disk of radius $R$, and
zero outside. A primitive cell of the honeycomb lattice, with
primitive vectors $\vec{a}_1$  and $\vec{a}_2$, is delimited in the
figure by the dotted lines. This primitive cell contains two
inequivalent elements each represented by a different colour. The
spacing between nearest neighbors is denoted by $a$. For the
numerical calculations it is convenient to consider as well a
rectangular supercell containing four elements (yellow region). The
primitive vectors of the reciprocal lattice, $\vec{b}_1$ and
$\vec{b}_2$, are represented in Fig. \ref{fig:honeycomb_lattice}
(right) together with the first Brillouin zone and with some
relevant high-symmetry points.

The microscopic Hamiltonian for this two-dimensional system is
simply
\begin{equation}
\Ham = \frac{-\hbar^2}{2m_b}\nabla^2+V(\vec{r}),
\label{E:Hamiltonian_honeycomb_lattice}
\end{equation}
where the electron effective mass $m_b$ is taken as in Ref. \cite{gibertini_engineering_2009}: $m_b=0.067m$, with $m$ the
electron rest mass.

As outlined in Sect. \ref{sec:homogenisation_scheme}, the first step
to compute the effective Hamiltonian $H_\text{ef}(\k,\omega)$ is to
find the microscopic wave function $\psi(\vec{r},\omega)$ that
satisfies Eq. \eqref{E:schrodinger_laplace} for a macroscopic
initial state $-i \hbar\psi_{t=0}(\vec{r}) = f_0 \e{i\k\cdot
\vec{r}}$ where $f_0$ is an arbitrary constant (the effective
Hamiltonian is independent of the value of $f_0$). Equation
\eqref{E:schrodinger_laplace} is numerically solved for a fixed
$(\k,\omega)$ using the finite-difference frequency domain (FDFD)
method. The details can be found in Appendix
\ref{sec:FDFD_honeycomb_lattice}. From the knowledge of
$\psi(\vec{r},\omega)$, the effective Hamiltonian
$H_\text{ef}(\k,\omega)$ is deduced as follows. Substituting the
identity $\left(\Ham_\text{ef} \Psi\right)(\vec{r},\omega) =
H_\text{ef}(\k,\omega) \Psi(\vec{r},\omega)$ and Eq.
\eqref{E:average_psi_1} into  Eq.
\eqref{E:averaged_Laplace_equation} it follows that:
\begin{equation}
 (H_\text{ef}(\k,\omega)-E) \psi_\text{av}(\k,\omega)\e{i\k\cdot \vec{r}}  = f_0 \cdot \e{i\k\cdot \vec{r}} ,
\end{equation}
where $\psi_\text{av}(\k,\omega)$ is defined as in Eq.
\eqref{E:average_psi_time} where $V_c$ should be understood as the
area of the unit cell. Thus, the single component effective
Hamiltonian in the spectral domain may be expressed as:
\begin{equation}
 H_\text{ef}(\k,\omega)  =  f_0 \cdot \psi_\text{av}^{-1}(\k,\omega)+E.
 \label{E:scalar_effective_hamiltonian_honeycomb_lattice}
\end{equation}

To give an example, we suppose that the structural parameters are
such that $V_0=-0.8~\milli\electronvolt$, $R/a=0.35$, and $a=150nm$.
The electronic band structure for this system was reported in Ref.
\cite{gibertini_engineering_2009}, and hence it will not be repeated
here.  As discussed in Ref. \cite{gibertini_engineering_2009}, for
an attractive potential it is possible to have an electronic band
structure with isolated Dirac points. For the chosen structural
parameters the tip of the Dirac cone at the $K$-point occurs at the
energy level $E_D = -0.326~\milli\electronvolt $. Because we are
interested in the physics near the $K$-point, it is implicitly
assumed that BZ in Eq. \eqref{E:BZ} represents the translation of
the first Brillouin zone to the $K$ point.

\begin{figure}[!h]
    \centering
 \includegraphics[width=0.6\linewidth]{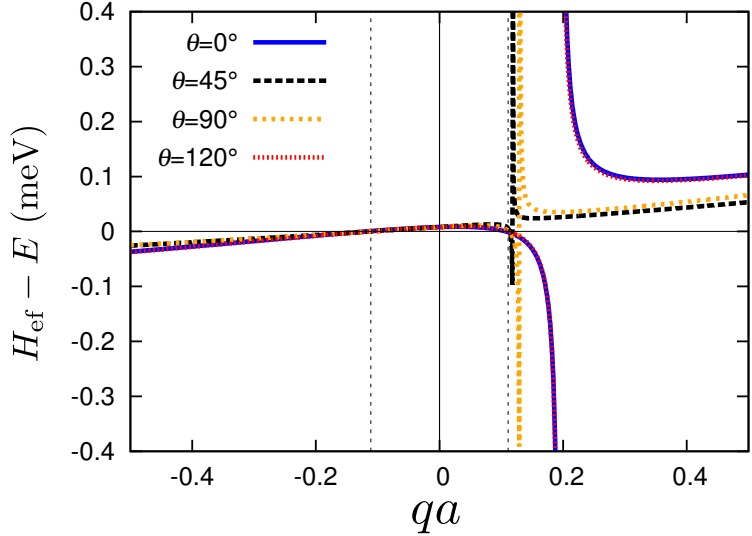}
          \caption{Plot of the single component effective Hamiltonians $H_\text{ef}$ near the $K$-point
           as a function of the normalized wave vector for different directions of propagation and $E=-0.33\,\milli\electronvolt, \,
            N_x=97, \, N_y=117, \, V_0=-0.8\,\milli\electronvolt$ and $R/a=0.35$. The vertical dashed lines indicate the zeros of $H_\text{ef}-E$.}
\label{fig:comparison_single_pseudospinor}
\end{figure}

In Fig. \ref{fig:comparison_single_pseudospinor} we depict the
calculated scalar effective Hamiltonian $H_\text{ef}$ for the energy
$E=-0.33~\milli\electronvolt$. In the numerical simulations we used
a grid with $N_x \times  N_y=97 \times 117$ points. The effective
Hamiltonian $H_\text{ef}$ is represented as a function of the wave
vector measured with respect to the $K$-point, so that $\vec{q}= \k
- \vec{K} $. The function actually plotted in the figure is
$H_\text{ef} - E$ versus the magnitude of the normalized wavevector
for different directions of propagation. The angle $\theta$ is the
angle between $\bf{q}$ and the $k_x$ axis (see Fig.
\ref{fig:honeycomb_lattice}).

As seen, for every direction $\theta$, the function $H_\text{ef} -
E$ intersects the horizontal axis at exactly one point (here,
negative values of $q$ are understood as being calculated for the
opposite direction $\theta + \pi$). Moreover, the position of the
zeros is nearly independent of the angle $\theta$. This is
consistent with the fact that the solutions of Eq.
\eqref{E:energy_states} give the dispersion of the stationary
states, which for a graphene-like system, due to the isotropy,
should depend only on $q$ and not on $\theta$. Importantly, Fig.
\ref{fig:comparison_single_pseudospinor} also demonstrates that the
effective Hamiltonian $H_\text{ef}$ depends strongly on $\vec{q}$,
and that it can have pole singularities for some specific values
$\vec{q}$. This implies that the time evolution operator is strongly
spatially dispersive. In other words, the inverse Fourier transform
in $\k$ of $H_\text{ef}(\k,\omega)$ is spread over a wide region of
space, which implies that the action of the effective Hamiltonian on
$\Psi$ is nonlocal. This property is undesired because the
associated formalism is cumbersome and lacks elegance. In the next
section, we prove that by considering a modified effective medium
approach wherein the averaged wave function is described by a
pseudospinor it is possible to overcome these problems.

\section{Two-component Hamiltonian for a potential with the honeycomb symmetry} \label{sec:pseudospinor}

The nonlocal spatial-action of the single component effective
Hamiltonian can be traced back to the fact that the electron wave
function can have significant fluctuations within each unit cell due
to the fact that the system is formed by two inequivalent
sublattices, i.e. that there are two inequivalent elements per unit
cell. This observation suggests that our definition of macroscopic
state may be too restrictive for this system, because it does not
allow us to consider electronic states that are more localized than the
unit cell, and hence the two sublattices are not discriminated.

As demonstrated in the following, it is possible to avoid these
problems by extending the definition of macroscopic states.

\subsection{Two components effective Hamiltonian} \label{Sec:subsect_pseudospinor}

\begin{figure}[!h]
    \centering
 \includegraphics[width=0.25\linewidth]{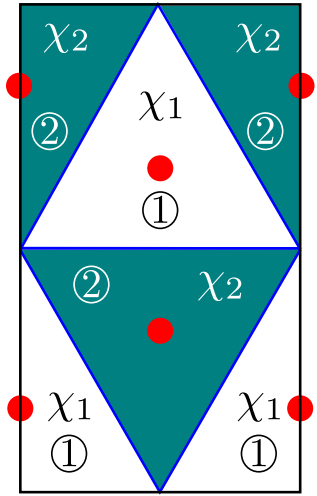}
          \caption{Representation of the $\chi_i(\vec{r})$ function that generates the pseudospinor wave function $\Phi$.}
\label{fig:chi_function}
\end{figure}

To begin with, let us decompose the crystal into two regions, each
described by a characteristic function $\chi_i(\vec{r})$ ($i$=1,2)
such that $\chi_1(\vec{r})+\chi_2(\vec{r})=1$. Specifically, the
characteristic functions are chosen such that $\chi_i(\vec{r})$
delimits a triangle centered on each disk of the same type and is
equal to 1 in the region
$\raisebox{.5pt}{\textcircled{\raisebox{-.9pt} {i}}}$ and 0 in the
complementary region (see Fig. \ref{fig:chi_function} and compare
with the supercell represented in Fig. \ref{fig:honeycomb_lattice}).
Note that this partition can be obtained through a process similar
to the one used to construct the Wigner-Seitz cell, but rather than
picking neighbor lattice points one picks elements from different
sublattices. Hence, $\chi_i(\vec{r})$ are the characteristic
functions associated with the two sublattices of the crystal.

Based on this decomposition, we introduce the notion of
\emph{generalized} macroscopic state, as a state that can be
decomposed as $\psi(\vec{r}) =\chi_1 (\vec{r}) \psi_1(\vec{r}) +
\chi_2 (\vec{r}) \psi_2(\vec{r})$ for some functions $\psi_i$
($i=1,2$) with $\psi_i(\vec{r}) =\left\{ \psi_i(\vec{r})
\right\}_\text{av}$. Clearly, this definition generalizes that of
Sect. \ref{sec:homogenisation_scheme}. The idea is now to introduce
a generalized effective Hamiltonian that allows us to characterize
the time evolution of generalized macroscopic states.

To this end, we introduce a pseudospinor given by
\begin{equation}
 \Phi= \begin{pmatrix} \Phi_1 \\ \Phi_2  \end{pmatrix}
 =  \begin{pmatrix} \psi\chi_1 \\ \psi\chi_2  \end{pmatrix}.
 \label{E:microscopic_pseudospinor}
\end{equation}
We note that $\psi= \Phi_1+\Phi_2$, and thus from the microscopic
Schr\"{o}dinger equation \eqref{E:Schrodinger_equation} it is
possible to write
\begin{equation}
(\chi_1+\chi_2)(\Ham-i \hbar \frac{\partial }{\partial
t})(\Phi_1+\Phi_2) = 0,
\end{equation}
This scalar equation is equivalent to the matrix system
\begin{equation}
\underbrace{  \begin{pmatrix} \chi_1 \Ham \chi_1 & \chi_1\Ham \chi_2 \\
\chi_2\Ham \chi_1 & \chi_2\Ham \chi_2  \end{pmatrix}}_{\Ham_g}
 \cdot \Phi= i \hbar \frac{\partial }{\partial t} \Phi,
 \label{E:hamiltonian_spinor}
\end{equation}
where $\Ham_g$ is a generalized two-component microscopic
Hamiltonian. We want to obtain an effective medium description of
the above time evolution problem with the macroscopic state given by
the spatially averaged pseudospinor $\Phi$.

To this end, assuming an initial generalized macroscopic state of
the form $\psi_{t=0}(\vec{r}) =\chi_1 (\vec{r})
\psi_{1,t=0}(\vec{r}) + \chi_2 (\vec{r}) \psi_{2,t=0}(\vec{r})$, we
calculate the Laplace transform of Eq. \eqref{E:hamiltonian_spinor},
to obtain:
\begin{equation}
\left( {\hat H_g  - E} \right) \cdot \Phi \left( {{\bf{r}},\omega }
\right) = \left( {\begin{array}{*{20}c}
   { - i\hbar \, \psi _{1,t = 0} \, \chi _1 }  \\
    - i\hbar \, \psi _{2,t = 0} \, \chi _2  \\
\end{array}} \right) \label{E:Laplace_generalized}
\end{equation}
The effective Hamiltonian operator $\hat H_{g,\text{ef}}$ is defined
so that $ \left( \Ham_{g,\text{ef}} \left\{\Phi \right\}_\text{av}
\right)(\vec{r},t) = \left\{ \left( \Ham_g \Phi \right)(\vec{r},t)
\right\}_\text{av}$ for any initial generalized macroscopic state
(compare with Eq. \ref{E:definition_effective_hamiltonian}). To find
an explicit formula for $\Ham_{g,\text{ef}} (\k, \omega)$ in the
Fourier domain we follow the same steps as in Sect.
\ref{sec:homogenisation_scheme}. Assuming that the initial state is
such that $-i \hbar \, \psi_{i,t=0}(\vec{r}) = f_i
\e{i\k\cdot\vec{r}}$ ($i=1,2$), with the weights $f_i$ generic
constants, it is simple to prove that:
\begin{equation}
\left( H_{g,\text{ef}}(\k,\omega)-E \right) \cdot \left\{ \Phi
\right\}_\text{av} = f_1 \left\{ \begin{pmatrix} \chi_1 \\ 0
\end{pmatrix} \e{i\k\cdot\vec{r}} \right\}_\text{av}
 +
 f_2 \left\{ \begin{pmatrix} 0 \\ \chi_2  \end{pmatrix} \e{i\k\cdot\vec{r}} \right\}_\text{av},
 \label{E:effective_hamiltonian_spinor_1}
\end{equation}
where $\left\{ \Phi \right\}_\text{av}$ is defined as
\begin{align}
 \left\{ \Phi \right\}_\text{av} &= \left( \frac{1}{V_c} \int \Phi(\vec{r},\omega) \cdot \e{-i\k\cdot\vec{r}} \Dr[N]{\vec{r}} \right) \e{i\k\cdot\vec{r}} \nonumber \\
 &= \Phi_\text{av} \e{i\k\cdot\vec{r}}.
 \label{E:psi_average_pseudospinor}
\end{align}
Because the
volume fraction of the two regions is identical, we have
\begin{equation}
 \left\{ \begin{pmatrix} \chi_1 \\ 0 \end{pmatrix} \e{i\k\cdot\vec{r}} \right\}_\text{av}
= \left( \frac{1}{V_c} \int \chi_1(\vec{r}) \Dr[N]{\vec{r}}  \right)
\begin{pmatrix} 1 \\ 0  \end{pmatrix} \e{i\k\cdot\vec{r}}
 =  \frac{1}{2} \begin{pmatrix} 1 \\ 0  \end{pmatrix}
 \e{i\k\cdot\vec{r}}.
\end{equation}
Thus, Eq. \eqref{E:effective_hamiltonian_spinor_1} becomes
\begin{equation}
\left[ H_{g,\text{ef}}(\k,\omega)-E \right] \cdot  \Phi_\text{av} =
f_1 \frac{1}{2} \begin{pmatrix} 1 \\ 0  \end{pmatrix}
 +
 f_2 \frac{1}{2} \begin{pmatrix} 0 \\ 1  \end{pmatrix}.
 \label{E:effective_hamiltonian_spinor}
\end{equation}
Let now $\Phi^{(1)}$ and $\Phi^{(2)}$ be the two independent
solutions of \eqref{E:Laplace_generalized}, corresponding
respectively to  $f_1=1$, $f_2=0$ and $f_1=0$, $f_2=1$. Then, it is
possible to write the matrix equation
\begin{equation}
\left[ H_{g,\text{ef}}(\k,\omega)-E \right] \cdot \begin{pmatrix}
\Phi_\text{av}^{(1)} & \Phi_\text{av}^{(2)}  \end{pmatrix} =
 \frac{1}{2} \begin{pmatrix} 1 & 0\\ 0 & 1 \end{pmatrix} ,
\end{equation}
or equivalently
\begin{equation}
 H_{g,\text{ef}}(\k,\omega)  = E +  \frac{1}{2} \begin{pmatrix} \Phi_\text{av}^{(1)} & \Phi_\text{av}^{(2)}  \end{pmatrix}^{-1}.
 \label{E:effective_hamiltonian_spinor_final}
\end{equation}
In summary, we demonstrated that the two component effective
Hamiltonian can be written in terms of the functions $\Phi^{(i)}$
($i$=1,2) as shown above.  From the definition, $\Phi^{(i)}$
satisfies Eq. \eqref{E:Laplace_generalized} with $- i\hbar \, \psi
_{j,t = 0} = \delta_{i,j} \e{i\k\cdot\vec{r}}$ ($j=1,2$). It is easy
to check that it can be written as
\begin{equation}
\Phi^{(i)} = \begin{pmatrix} \psi^{(i)} \chi_1 \\
\psi^{(i)} \chi_2 \end{pmatrix}
\end{equation}
where $\psi^{(i)}$ ($i$=1,2) is the solution of the scalar problem
\begin{equation}
(\Ham-E)\psi^{(i)}=\chi_i\e{i\k\cdot\vec{r}}.
\end{equation}
Thus, to compute the two-component Hamiltonian
$H_{g,\text{ef}}(\k,\omega)$ one needs to solve two independent
scalar problems. This operator describes exactly the time evolution
of any generalized macroscopic state with a pseudospinor formalism.

\subsection{Single component Hamiltonian obtained from the pseudospinor
formalism} \label{sec:single_component_Hamiltoninans_honeycomb}

It should be clear from the previous subsection that
$H_{g,\text{ef}}$ is a generalization of $H_{\text{ef}}$ defined in
Sect. \ref{sec:honeycomb_lattice}. This suggests that
$H_{\text{ef}}$ can be written in terms of $H_{g,\text{ef}}$. In the
following, we prove that this is the case.

To begin with, we note that solving Eq. \eqref{E:schrodinger_laplace} for a macroscopic
initial state $-i \hbar\psi_{t=0}(\vec{r}) = f_0 \e{i\k\cdot \vec{r}}$ is equivalent to solve Eq. \eqref{E:Laplace_generalized} with an initial state
 $-i \hbar \, \psi_{i,t=0}(\vec{r}) = f_i
\e{i\k\cdot\vec{r}}$ ($i=1,2$) with $f_1=f_2=f_0$. Fixing $f_1=f_2=f_0$ in Eq. \eqref{E:effective_hamiltonian_spinor},
we find after straightforward manipulations that for such an excitation
\begin{equation}
 \Phi_\text{av} =  \frac{f_0}{2} \left[ H_{g,\text{ef}}(\k,\omega)-E \right]^{-1} \cdot  \begin{pmatrix} 1 \\ 1  \end{pmatrix} .
\end{equation}
Then, noting that $ \psi_\text{av}= \Phi_{1,\text{av}}+\Phi_{2,\text{av}} $, it follows that
\begin{equation}
 \psi_\text{av} =  \frac{f_0}{2} \sum_{i,j} \left[  \left( H_{g,\text{ef}}(\k,\omega)-E \right)^{-1} \right]_{i,j},
\end{equation}
and thus, by substitution into equation
\eqref{E:scalar_effective_hamiltonian_honeycomb_lattice}, the single
component Hamiltonian $\Ham_\text{ef}^\text{(s)}$ deduced from
$\Ham_{g,\text{ef}}$ is
\begin{equation}
H_\text{ef}^\text{(s)}(\k,\omega)   =   \frac{2}{\displaystyle \sum_{i,j} \left[  \left( H_{g,\text{ef}}(\k,\omega)-E \right)^{-1} \right]_{i,j}} + E.
\label{E:relation_psiAv_Hef}
\end{equation}
We numerically verified that $H_\text{ef}^\text{(s)}$ given by the
above formula is exactly coincident with scalar effective
Hamiltonian, $\Ham_\text{ef}$, obtained with the calculation method
described in Sect. \ref{sec:honeycomb_lattice}, as it should be.
This coincidence supports the correctness of our numerical codes.

It is interesting to note that Eq. \eqref{E:relation_psiAv_Hef}
explicitly shows that the zeros of $H_\text{ef}^\text{(s)}(\k,\omega) -
E$ occur for the same values of $(\k,\omega)$ as the poles of $\left(
H_{g,\text{ef}}(\k,\omega)-E \right)^{-1}$. This is equivalent to say
that the zeros of $H_\text{ef}(\k,\omega) - E$ are coincident with the
zeros of $\det\left( H_{g,\text{ef}}(\k,\omega)-E \right)$. This is a
consequence of the fact that the electronic band structure of the
microscopic Hamiltonian is exactly predicted by the two effective
medium formulations \cite{silveirinha_effective_2012}.

\subsection{Stationary states near the $K$-point}

As mentioned in section \ref{sec:homogenisation_scheme}, the
stationary electronic states of the 2DEG can be obtained from the
pseudospinor effective Hamiltonian $H_{g,\text{ef}}$ by finding the
solutions of:
\begin{equation}
\det \left( H_{g,\text{ef}}(\k,E)-E \right)  = 0.
\label{E:determinant_pseudo_spinor}
\end{equation}
In this work, we are mainly interested in the physics near the high-symmetry
$\vec{K}$ point. Hence, it is convenient to simplify
the formalism and use an analytical approximation for
$H_{g,\text{ef}}$ to solve the secular equation. For
electron states with the spatial spectrum concentrated near the
$\vec{K}$ point, we can approximate $H_{g,\text{ef}}$ by its Taylor
series:
\begin{equation}
H_{g,\text{ef}}(\k,E) \simeq H_{g,\text{ef}}(\vec{K},E) + \left.
\frac{\partial H_{g,\text{ef}}(\k,E)}{\partial
k_x}\right|_{\vec{k}=\vec{K}} q_x +\left. \frac{\partial
H_{g,\text{ef}}(\k,E)}{\partial k_y}\right|_{\vec{k}=\vec{K}} q_y,
\label{E:Taylor_Hef_honeycomb}
\end{equation}
with $k_x$,  $k_y$, and $q_x$, $q_y$ the components of the wave
vectors $\k$ and $\vec{q}=\vec{k}-\vec{K}$, respectively. In
practice, the derivatives of the effective Hamiltonian are
numerically evaluated with finite differences. The band energy
diagram $E(\k)$ of the system is then obtained by solving
\eqref{E:determinant_pseudo_spinor} using the approximate expression
of $H_{g,\text{ef}}(\k,E)$. Notably, we numerically verified that
the two component Hamiltonian is a smooth slowly-varying function of
$\vec{q}$ (not shown), and hence the above Taylor expansion is
typically a quite good approximation for the Hamiltonian. This
contrasts with the singular behavior of the single component
effective Hamiltonian (see Fig.
\ref{fig:comparison_single_pseudospinor}).

\begin{figure}[!h]
    \centering
 \includegraphics[width=\linewidth]{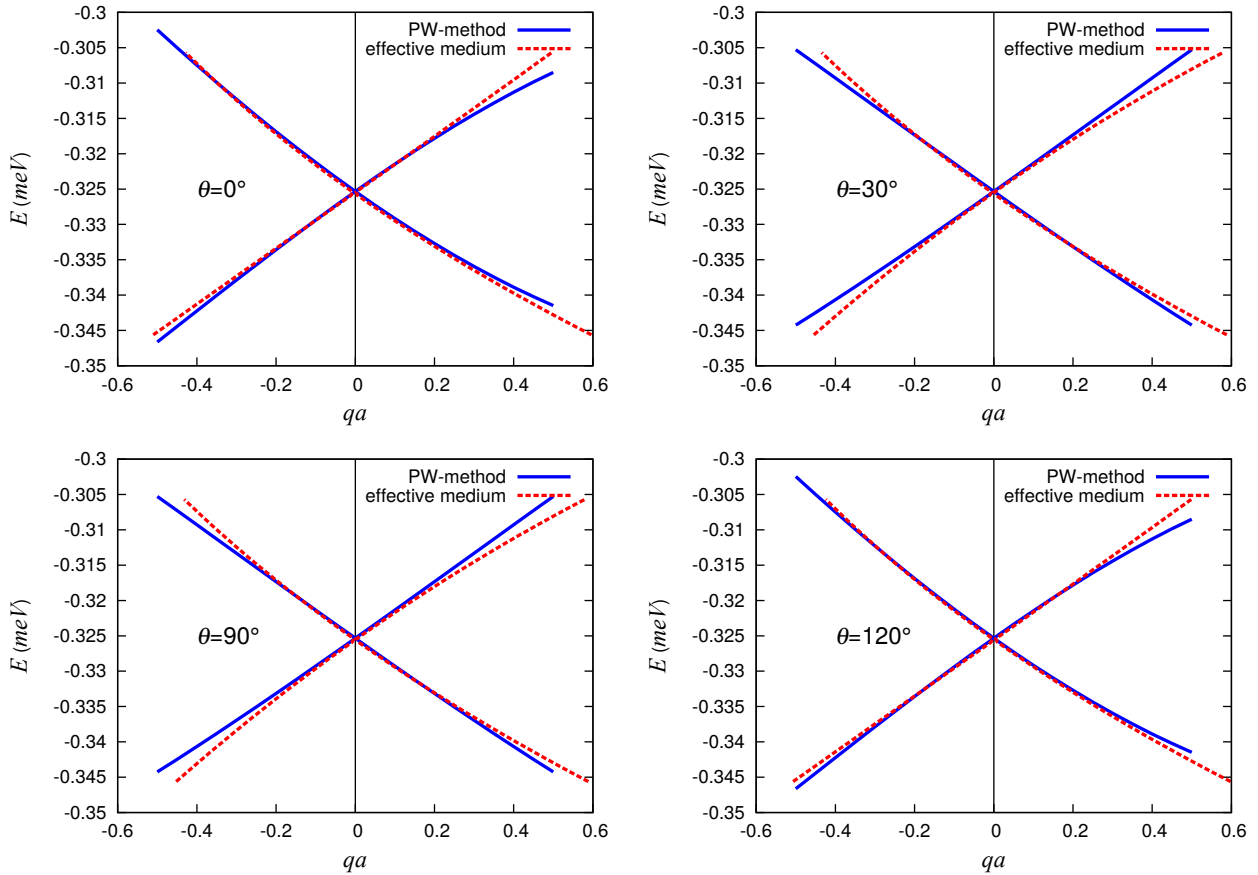}
          \caption{Electronic band structure near the $K$-point for different propagation directions and for a modulated 2DEG
    with a potential $V_0=-0.8~\milli\electronvolt$, a ratio $R/a=0.35$, and a number of nodes in the FDFD grid $N_x=97$ and $N_y=117$.
    Dashed red lines: calculated with the two-component effective Hamiltonian. Solid blue lines: calculated with the plane wave method.}
\label{fig:band_diagram_pseudospinors}
\end{figure}

The energy dispersion diagrams obtained with this approach for a
system with the same parameters as in Fig.
\ref{fig:comparison_single_pseudospinor} ($R/a=0.35$, and
$V_0=-0.8~\milli\electronvolt$) are depicted in Fig.
\ref{fig:band_diagram_pseudospinors} (dashed lines). Each plot
corresponds to a specific angle of propagation $\theta$, with
$\theta$ defined in the same way as in Fig.
\ref{fig:comparison_single_pseudospinor}.
The solid lines correspond to the exact electronic band structure,
and were obtained with the plane wave method
\cite{ashcroft_solid_1976,joannopoulos_photonic_2008} by solving
\begin{equation}
\det \left[\left( \frac{\hbar^2}{2m}(\k-\vec{G})^2 - E_n(\k)\right)
\delta_{\vec{G},\vec{G}^\prime} + \tilde{V}(\vec{G}^\prime-\vec{G})
\right] = 0
\end{equation}
where $\vec{G}$, $\vec{G}^\prime$ are the reciprocal lattice vectors
and $\tilde{V}$ is the Fourier transform of the potential.
As seen in Fig. \ref{fig:band_diagram_pseudospinors}, there is a
very good agreement between the plane-wave and effective medium
results near the $K$-point for all the directions of propagation.
Note that the results are not exactly coincident because of the
approximation implicit in the Taylor expansion of $H_{g,\text{ef}}$
near the $K$-point. However, the proximity between the two sets of
curves confirms that Eq. \eqref{E:Taylor_Hef_honeycomb} is, indeed,
quite accurate. Moreover, our effective medium results corroborate
the findings of Ref. \cite{gibertini_engineering_2009}: near the
$K$-point, the energy dispersion diagrams are linear, isotropic,
with a zero band-gap. These properties are not the only similarities
of the modulated 2DEG with graphene. In fact, we shall prove in the
following that the electronic states pseudospinor may also be
determined by a 2D massless Dirac fermion Hamiltonian $\Ham_D=\hbar
v_F \boldsymbol{\sigma} \cdot \bf{q}$, where $v_F$ is the equivalent
``Fermi velocity'' and $\boldsymbol{\sigma}$ are the Pauli matrices.
However, in order to do this we will need to renormalize the
pseudospinor. The reason is discussed in the next subsections.

\subsection{Macroscopic Probability Density for the Stationary States}

The probability density is of fundamental importance in
quantum mechanics since it is essential to make physical
predictions. Within the usual microscopic framework, it is given by
$\mathcal{P}_\text{mic}= \psi^\ast \cdot \psi$. Evidently, it can
also be written in terms of the pseudospinor
\eqref{E:microscopic_pseudospinor} as $\mathcal{P}_\text{mic}=
\Phi^\ast \cdot \Phi$. Hence, the average probability density for a
Bloch wave is:
\begin{equation}
 \mathcal{P}_\text{mic,av}=\left\{ \Phi^\ast
\cdot \Phi \right\}_\text{av} = \frac{1}{V_c}\int \Phi^\ast \cdot
\Phi \, \Dr[N]{\vec{r}}.
\end{equation}
One important observation is that, in general, the averaging operation
does not commute with multiplication operation:
\begin{equation}
\left\{ \Phi^\ast \cdot \Phi \right\}_\text{av}  \neq \left\{ \Phi
\right\}_\text{av}^\ast \cdot \left\{ \Phi \right\}_\text{av}.
\end{equation}
This indicates that in general the squared amplitude of the
spatially-averaged wave function cannot be identified with the
probability density in the macroscopic framework. This may look
peculiar at first sight, but actually the situation is quite
analogous to what happens in macroscopic electrodynamics wherein the
formula for the stored energy calculated using the macroscopic
electromagnetic fields differs from the formula for the stored
energy calculated using the microscopic fields \cite{Landau,
silveirinha_poynting_2009}.

It is demonstrated in Appendix
\ref{sec:Probability_density_pseudospinor} (see also the
supplementary materials of Ref. \cite{fernandes_wormhole}) that for
stationary (Bloch) electronic states the following relation holds
exactly
\begin{equation}
\label{E:density_probability_average_psi} \left\{ \Phi^\ast \cdot
\Phi \right\}_\text{av} = 2 \left\{ \Phi \right\}_\text{av}^\ast
\cdot \left( {\bf{1}} -  \frac{\partial
\Ham_{g,\text{ef}}(\k,E)}{\partial E}\right) \cdot \left\{ \Phi
\right\}_\text{av}.
\end{equation}
Hence, we can write
\begin{equation}
\mathcal{P}_\text{mic,av} =\mathcal{P}_\text{mac},
\end{equation}
where the macroscopic probability density is defined as
\begin{equation}
\mathcal{P}_\text{mac} =2 \left\{ \Phi \right\}_\text{av}^\ast \cdot
\left( {\bf{1}} -  \frac{\partial \Ham_{g,\text{ef}}(\k,E)}{\partial
E}\right) \cdot \left\{ \Phi \right\}_\text{av}.
\end{equation}
Thus, when the effective Hamiltonian is energy dependent, i.e. in
presence of temporal dispersion, the formula for the macroscopic
probability density differs from that of the microscopic probability
density.

For convenience, we define the Dirac energy, $E_D$, as the energy
for which the valence and conduction bands coincide. Interestingly,
our numerical calculations indicate that near the Dirac energy
${\bf{1}} - \frac{\partial \Ham_{g,\text{ef}}(\k,E)}{\partial E}$
varies slowly. Thus, the macroscopic probability density may be
approximated by:
\begin{equation}
\mathcal{P}_\text{mac} \approx \mathcal{P}_\text{mac}^0 =2 \left\{
\Phi \right\}_\text{av}^\ast \cdot \vec{A}_0 \cdot \left\{ \Phi
\right\}_\text{av}, \label{E:Pmac}
\end{equation}
with $ \vec{A}_0 =  {\bf{1}} -  \left. \frac{\partial
\Ham_{g,\text{ef}}(\k,E)}{\partial E}\right|_{E=E_D,\vec{k}=\vec{K}}
$.
\begin{figure}[!h]
    \centering
 \includegraphics[width=0.65\linewidth]{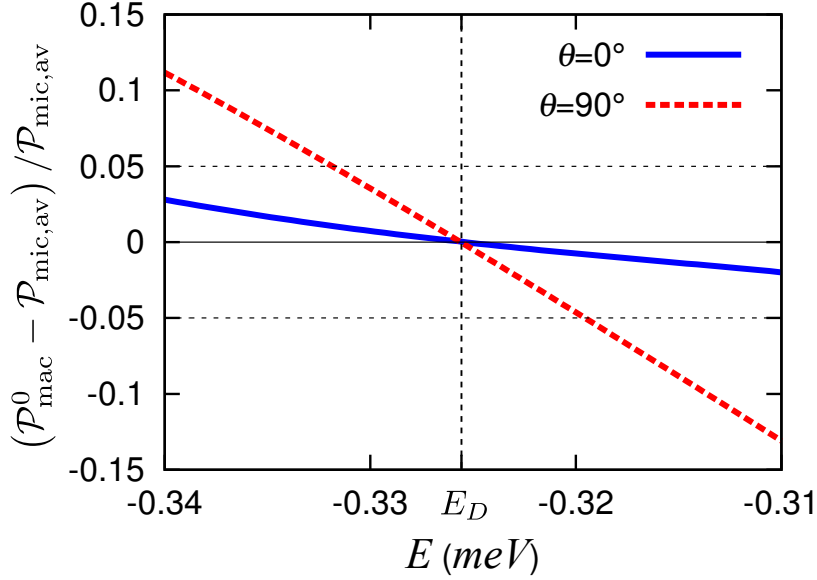}
         \caption{Relative difference between $ \mathcal{P}_\text{mic,av}$ and $ \mathcal{P}_\text{mac}^0$ as a function of energy for a modulated
         2DEG with the same parameters as in Fig.
         \ref{fig:band_diagram_pseudospinors}. Two different directions of propagation are considered.}
\label{fig:relative_error_density_probability}
\end{figure}

To  confirm the validity of this formula, we numerically calculated
the relative difference between $ \mathcal{P}_\text{mic,av}$ and $
\mathcal{P}_\text{mac}^0$ for the stationary states of the system
near $E_D$ and for two directions of propagation $\theta$. As seen
in Fig. \ref{fig:relative_error_density_probability}, in the
considered energy range the error in the approximation is smaller
than 15 \% for a direction of propagation with $\theta=90 \degree$,
and smaller than 5 \% for $\theta=0 \degree$. Notice that the error
vanishes at the Dirac energy because in this case
$\mathcal{P}_\text{mac}$ is equal to $\mathcal{P}_\text{mac}^0$, and
from Eq. \eqref{E:density_probability_average_psi}  the relative
difference between the probability densities is exactly zero at this
point. We also verified (not shown) that within numerical precision,
$ \mathcal{P}_\text{mic,av} = \mathcal{P}_\text{mac}$ for all energy
values.

\subsection{Massless Dirac equation} \label{sec:massless_Dirac_equation}

We are now ready to show that the modulated 2DEG may be described by the
massless Dirac equation. The starting point is to generalize Eq.
\eqref{E:Taylor_Hef_honeycomb} and expand the effective Hamiltonian
in the spectral domain in a Taylor series near $E_D$ and $\vec{K}$
so that
\begin{equation}
(\Ham_{g,\text{ef}} - E) \cdot \left\{ \Phi \right\}_\text{av}
\simeq -(E-E_D) \vec{A}_0 \cdot \left\{ \Phi \right\}_\text{av} +
q_x \vec{A}_1 \cdot \left\{ \Phi \right\}_\text{av}+ q_y \vec{A}_2
\cdot \left\{ \Phi \right\}_\text{av}, \label{E:taylor_matricial}
\end{equation}
where the matrix $\vec{A}_0$ is defined as in the previous
subsection, $\vec{A}_1 =  \left. \frac{\partial
\Ham_{g,\text{ef}}(\k,E)}{\partial
k_x}\right|_{E=E_D,\vec{k}=\vec{K}} $ and $ \vec{A}_2 =  \left.
\frac{\partial  \Ham_{g,\text{ef}}(\k,E)}{\partial
k_y}\right|_{E=E_D,\vec{k}=\vec{K}}$. Next, we introduce a
renormalized pseudospinor
\begin{equation}
\Phi_D= \sqrt{2} \cdot \vec{A}_0^{1/2} \cdot \left\{ \Phi
\right\}_{{\rm{av}}} e^{ - i{\bf{K}} \cdot {\bf{r}}},
\label{E:phi_D}
\end{equation}
which from Eq. \eqref{E:Pmac} is such that for stationary states the
probability density is given by the squared amplitude of the
renormalized pseudospinor $ \mathcal{P}_\text{mac} \approx \Phi_D
\cdot \Phi_D^\ast$. Note that the $\vec{A}_0$ matrix is necessarily
positive definite and is not unitary. The secular equation $(\Ham_{g,\text{ef}} - E)
\cdot \left\{ \Phi \right\}_\text{av} = 0$ is equivalent to
$(\Ham_{D} - E) \cdot \Phi_D = 0$ with $(\Ham_{D} - E) =
\vec{A}_0^{-1/2} \cdot (\Ham_{g,\text{ef}} - E) \cdot
\vec{A}_0^{-1/2}$. Simple manipulations show in the spatial domain:
\begin{equation}
\hat H_D  = E_D   - i\left( {\frac{\partial }{{\partial
x}}{\bf{\tilde A}}_1  + \frac{\partial }{{\partial y}}{\bf{\tilde
A}}_2 } \right)
\end{equation}
where $\tilde{\vec{A}}_1 = \vec{A}_0^{-1/2} \cdot \vec{A}_1 \cdot
\vec{A}_0^{-1/2}$ and $\tilde{\vec{A}}_2 = \vec{A}_0^{-1/2} \cdot
\vec{A}_2  \cdot \vec{A}_0^{-1/2}$. Interestingly, our numerical
calculations reveal (see Appendix \ref{sec:ApC}) that
$\tilde{\vec{A}}_1$ and $\tilde{\vec{A}}_2$ are of the form
\begin{gather}
 \tilde{\vec{A}}_1=\hbar v_F( \cos \phi \cdot \sigma_x - \sin \phi \cdot \sigma_y) , \nonumber\\
 \tilde{\vec{A}}_2=\hbar v_F(\sin \phi \cdot \sigma_x + \cos \phi \cdot \sigma_y) , \label{E:Atil}
\end{gather}
where $\phi \approx 60\degree$, $v_F$ is some constant that depends on the structural parameters of the 2DEG, and $\sigma_x= \begin{pmatrix} 0 & 1\\
1& 0\end{pmatrix}$ and $\sigma_y= \begin{pmatrix} 0 & -i\\ i&
0\end{pmatrix}$ are the usual Pauli matrices. Thus, the operator
$\hat H_D$ can be written in a compact form as:
\begin{equation}
\hat H_D  = E_D   - i \hbar v_F \left( {\frac{\partial }{{\partial
x'}} \sigma_x  + \frac{\partial }{{\partial y'}} \sigma_y } \right),
\end{equation}
where $\frac{\partial }{{\partial x'}}$ and $\frac{\partial
}{{\partial y'}}$ are the directional derivatives along the
directions $\phi = 60\degree$ and $\phi = 60\degree + 90\degree$,
respectively:
\begin{equation}
\left\{ \begin{array}{l}
 \frac{\partial }{{\partial x'}} = \cos \phi \frac{\partial }{{\partial x}} + \sin \phi \frac{\partial }{{\partial y}} \\
 \frac{\partial }{{\partial y'}} =  - \sin \phi \frac{\partial }{{\partial x}} + \cos \phi \frac{\partial }{{\partial y}} \\
 \end{array} \right.
\end{equation}
Thus, $\hat H_D$ is exactly the 2D massless Dirac Hamiltonian, and
the pseudospinor associated with stationary states is a solution of
the time-independent Dirac equation $(\Ham_{D} - E) \cdot \Phi_D =
0$. It should be noted that the original coordinate axes need to be
rotated by $\phi = 60\degree$ to get an Hamiltonian operator
consistent with that of graphene. Our honeycomb lattice is actually rotated by $30\degree$ with respect to the definition usually
adopted for graphene \cite{castro_neto_electronic_2009}. It can be verified that after a suitable similarity transformation $\hat H_D$ assumes the usual form in the standard coordinate system of graphene.

We would like to underline that in order to obtain a 2D
massless Dirac Hamiltonian it was essential to renormalize the
pseudospinor such that for stationary states, $
\mathcal{P}_\text{mac} \approx \Phi_D \cdot \Phi_D^\ast$, because
only in these conditions the analogy with graphene is complete. Notably,
without this renormalization the Hamiltonian $\Ham_{g,\text{ef}}$ is not
equivalent to a massless 2D Dirac Hamiltonian.

To further explore the analogy with graphene, next we numerically
confirm that each component of the pseudospinor corresponds to a
state localized on a different sublattice of the 2DEG. The
eigenfunctions of the Dirac Hamiltonian $\Ham_{D}$ in the conduction
band are proportional to $\begin{pmatrix} 1 \\ \e{i\theta_\vec{q}}
\end{pmatrix}$, where  $\theta_\vec{q}=\phi +
\arctan\left(\frac{q_y}{q_x}\right)$ and $q_x$, $q_y$ are measured
relatively to the Dirac point. Note that $\theta_\vec{q}$ depends on
the rotation angle $\phi = 60 \degree$ previously discussed. Hence,
the two components of the pseudospinor $\Phi_D$ are in phase for
$\theta_\vec{q}=\phi$ and out of phase for $\theta_\vec{q}=\phi -
\pi$. To verify the connection between the microscopic and the
macroscopic theories we numerically calculated the microscopic wave
function associated with a wave vector $\vec{q}$ oriented along the
directions $\theta_\vec{q}=60 \degree$ and $\theta_\vec{q}=-120
\degree$. In the simulations it was assumed that $R/a=0.35$,
$V_0=-0.8~\milli\electronvolt$, and that $E-E_D = -0.329 ~\milli\electronvolt$.
\begin{figure}[!th]
    \centering
 \includegraphics[width=1\linewidth]{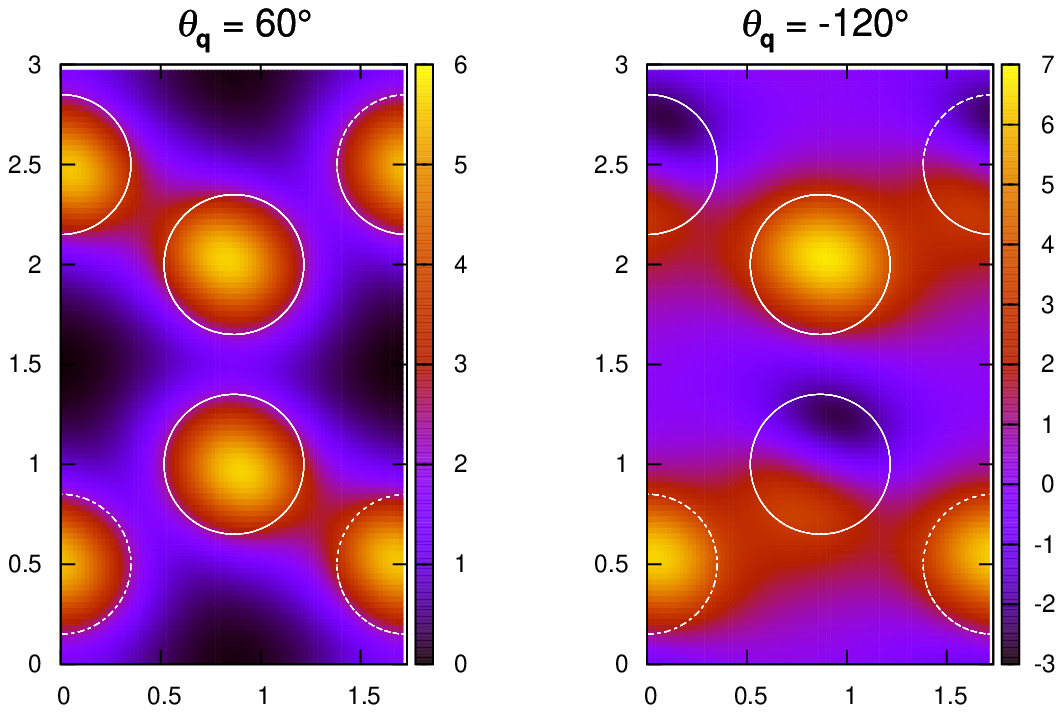}
         \caption{Density plot of
${\mathop{\rm Re}\nolimits} \left\{ {\frac{1}{{{\Phi _{1,{\rm{av}}}}}}\psi \left( {{\bf{r}},\omega } \right){e^{ - i{\bf{k}} \cdot {\bf{r}}}}} \right\}$
for $\theta_\vec{q}=60 \degree$ and $\theta_\vec{q}=-120 \degree$.}
\label{fig:thetaK}
\end{figure}

Figure \ref{fig:thetaK} depicts the numerically calculated functions
${\mathop{\rm Re}\nolimits} \left\{ {\frac{1}{{{\Phi
_{1,{\rm{av}}}}}}\psi \left( {{\bf{r}},\omega } \right){e^{ -
i{\bf{k}} \cdot {\bf{r}}}}} \right\}$, i.e. the real part of the
wave function envelope normalized to the first component $\Phi
_{1,{\rm{av}}}$ of the pseudospinor ${\left\{ \Phi
\right\}_{{\rm{av}}}}$. The normalization to $\Phi _{1,{\rm{av}}}$
is done to ensure that the argument of the ${\mathop{\rm
Re}\nolimits} \left\{ {} \right\}$ operator is dominantly
real-valued. The white circles in Fig. \ref{fig:thetaK} represent
the positions of the disks where the potential is applied. One can
see that for an angle of $\theta_\vec{q}=60 \degree$, the wave
function envelope is in phase inside all disks, and therefore both
components of the pseudospinor \eqref{E:microscopic_pseudospinor}
are also in phase. On the contrary, for an angle of
$\theta_\vec{q}=-120 \degree$, the wave function envelope is out
of phase inside the disks of the two different sublattices and thus
the same is true for the components of the pseudospinor, as we
wanted to show.

Up to now, the discussion was focused in the stationary states of
the modulated 2DEG. Notably, the operator $\hat H_D$ also describes
the time dynamics of generalized macroscopic states. This can be
easily demonstrated by calculating the inverse Laplace-Fourier
transform of the right-hand side of Eq. \eqref{E:taylor_matricial}
and noting that for a time evolution problem it must vanish for
$t>0$. This yields:
\begin{equation}
i\hbar \frac{\partial }{{\partial t}}{\bf{A}}_0  \cdot \tilde \Phi
_{{\rm{av}}}  = E_D {\bf{A}}_0  \cdot \tilde \Phi _{{\rm{av}}}  -
i\left( {\frac{\partial }{{\partial x}}{\bf{A}}_1  + \frac{\partial
}{{\partial y}}{\bf{A}}_2 } \right) \cdot \tilde \Phi _{{\rm{av}}},
\,\,\, t>0
\end{equation}
where $ \tilde \Phi _{{\rm{av}}}  = \left\{ \Phi
\right\}_{{\rm{av}}} e^{ - i{\bf{K}} \cdot {\bf{r}}}$ is the
envelope of the macroscopic wave function. Using now the definition
of $\Phi_D$ (Eq. \eqref{E:phi_D}) it is easy to show that:
\begin{equation}
i\hbar \frac{{\partial \Phi _D }}{{\partial t}} = \hat H_D  \cdot
\Phi _D,
\end{equation}
and hence the time evolution of generalized macroscopic states is
indeed described by the massless 2D Dirac equation.

\subsection{Parametric study} \label{sec:tunability_honeycomb_lattice}

\begin{figure}[!h]
    \centering
 \includegraphics[width=\linewidth]{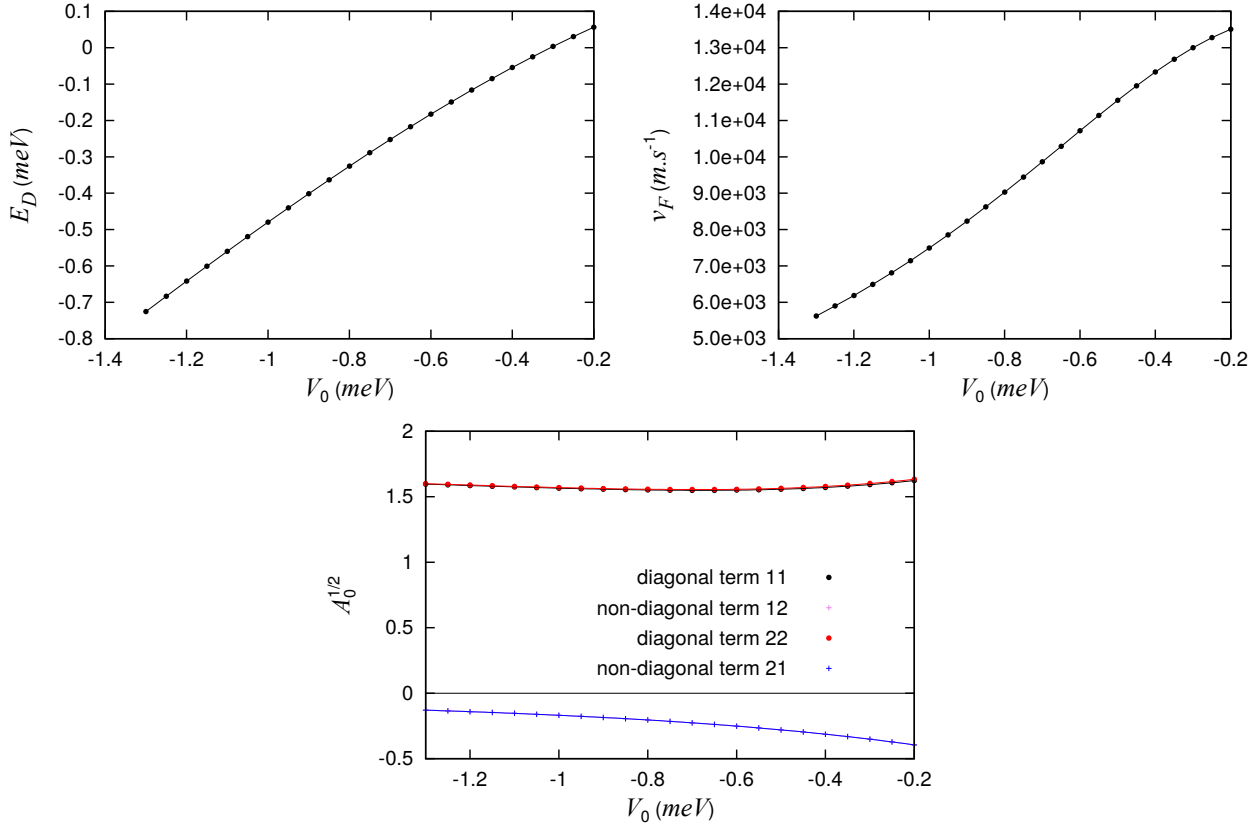}
          \caption{Dirac energy, Fermi velocity and elements of the $\vec{A}_0^{1/2}$ matrix as a function of the potential $V_0$
    obtained with the effective medium theory for the Dirac cone near the $K$-point.
    In these simulations it was assumed that $R/a = 0.35$.}
\label{fig:EF_vF_A0}
\end{figure}

\begin{figure}[!h]
    \centering
 \includegraphics[width=\linewidth]{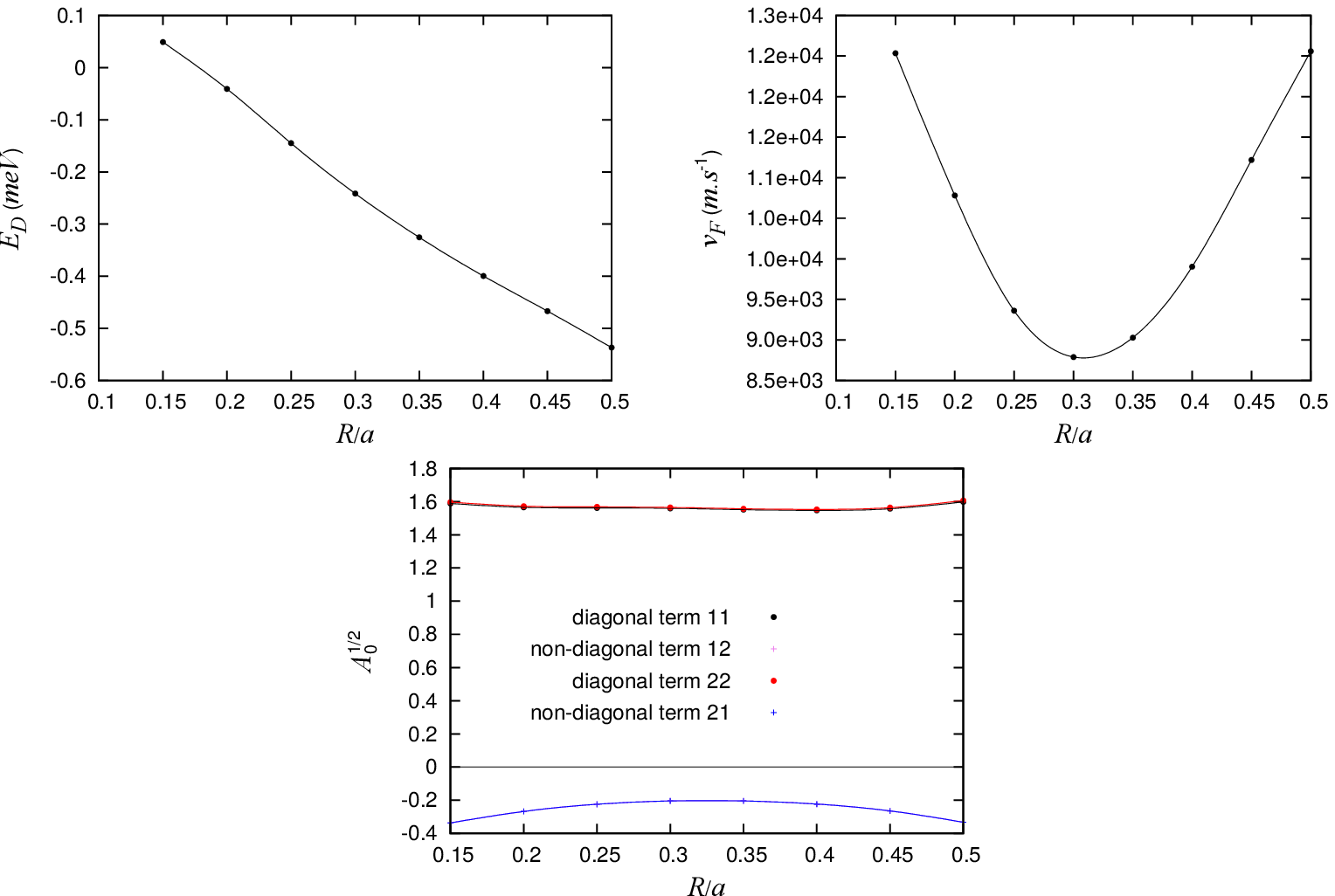}
          \caption{Dirac energy, Fermi velocity and elements of the $\vec{A}_0^{1/2}$ matrix as a function of
     $R/a$ obtained with the effective medium theory for the Dirac cone near the $K$-point.
     In these simulations it was assumed that $V_0 = -0.8~\milli\electronvolt$.}
\label{fig:EF_vF_A0_vs_r/a}
\end{figure}

By varying the geometric parameters or the strength of the potential
it is possible to tune the characteristics of the Dirac cones.
Hence, it is relevant to present a parametric study of the effective
Hamiltonian parameters. Figures \ref{fig:EF_vF_A0} and
\ref{fig:EF_vF_A0_vs_r/a} show the dependance of the Dirac energy
$E_D$, of the Fermi velocity $v_F$ and of the elements of the matrix
$\vec{A}_0^{1/2}$ with the strength of the potential $V_0$ and with
the normalized disk radius $R/a$, respectively. The range of values
considered for $V_0$ is such that the only available stationary
states near $E_D$ are associated with the Dirac cones, consistently
with the study of Ref. \cite{gibertini_engineering_2009}.

First, we remark that the matrix $\vec{A}_0^{1/2}$ is an (almost
diagonal) real-valued symmetric matrix whose elements remain almost constant
when changing either the potential or the normalized radius $R/a$.
Also, as expected, $E_D$ becomes more negative as $V_0$ is decreased
and as $R/a$ is increased.


On the other hand, consistent with what is reported in Ref.
\cite{gibertini_engineering_2009}, it is seen that the Fermi
velocity increases as the absolute value of the potential is
decreased, and exhibits a parabolic dependence on $R/a$. Moreover,
the value of $v_F$ is of the same order of magnitude as $v_F^{\left(
{nf} \right)}  = \frac{{2\pi \hbar }}{{3\sqrt 3 m_b a}} = 1.4 \cdot
10^4 \meter\second^{-1}\,\,$ \cite{gibertini_engineering_2009},
which is roughly two orders of magnitude smaller than in graphene.
Here, we would like to note that the value for $v_F^{\left( {nf}
\right)}$ reported in Ref. \cite{gibertini_engineering_2009} is
overestimated by a factor of 10, likely due to a typo. Linear
dispersing bands have exciting applications in terahertz photonics,
and in the enhancement of the nonlinear optical response
\cite{silveirinha_giant_2014,mikhailov_nonlinear_2008}.


\section{$\text{HgCdTe}$ hexagonal superlattice} \label{sec:triangular_superlattice}

In the second part of this article, we apply the effective medium
formalism to a different physical system with linearly dispersing
bands. Specifically, in a previous work
\cite{silveirinha_giant_2014} we have shown how by combining
mercury-cadmium-telluride (HgCdTe) semiconductor alloys it may be
possible to realize a superlattice \cite{Esaki} with an isotropic
zero-effective mass and a single valley linear energy-momentum
dispersion near the $\Gamma$-point. Here, we compute the effective
Hamiltonian of the superlattice, and demonstrate that in this second
platform the electrons do not have a pseudospin. HgCdTe quantum
wells have recently elicited great attention in the context of the
quantum spin Hall effect \cite{Bernevig, Konig}.
\begin{figure}[!h]
    \centering
 \includegraphics[width=0.5\linewidth]{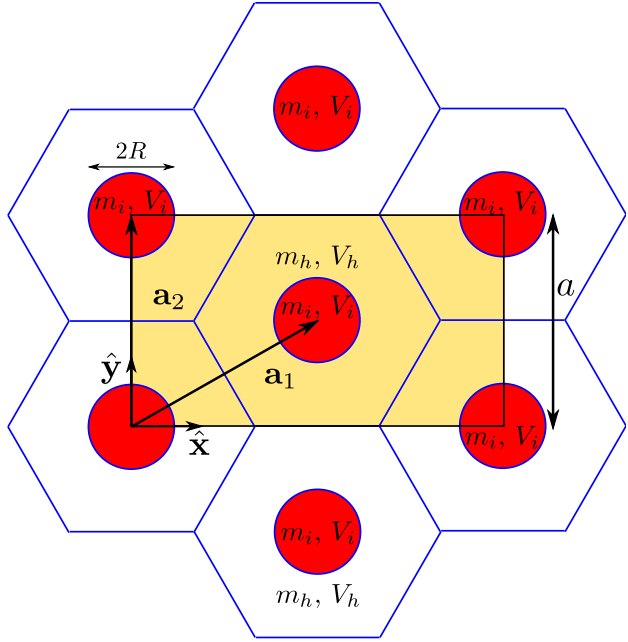}
         \caption{Hexagonal superlattice with primitive vectors $\vec{a}_1$ and $\vec{a}_2$.
         The potential and electron effective mass are $V_h$ and $m_h$ outside the disks, and $V_i$ and $m_i$ inside.
           The rectangular supercell used for the FDFD discretization is represented by the coloured area.}
\label{fig:superlattice}
\end{figure}

\subsection{Microscopic Hamiltonian}

The geometry of the heterostructure under study is depicted in Fig.
\ref{fig:superlattice}. Similar to the previous sections, it is a
two-dimensional structure (we are only interested in propagation in
the $xoy$ plane) formed by two lattice matched semiconductors. As
discussed in Ref. \cite{silveirinha_metamaterial-inspired_2012}, the
physics of electron waves in binary compounds with a zincblende-type
structure may be determined based on a potential $V$ and on a
dispersive (energy dependent) effective mass parameter $m$. In Fig.
\ref{fig:superlattice} the host material (in the exterior region) is
characterized by parameters $V_h$ and $m_h$, whereas the
``disk''-type inclusions are characterized by the parameters $V_i$
and $m_i$. As detailed below, $V$ and $m$ depend on the energy
levels of the conduction and valence bands of each material.
Consistent with the analysis of Refs. \cite{silveirinha_giant_2014,silveirinha_metamaterial-inspired_2012},
and with the generalized Ben Daniel-Duke boundary conditions
\cite{bastard_wave_1988,chuang_physics_1995,voon_k_2009}, this
heterostructure may be modeled by a Hamiltonian $\Ham$ such that
\begin{equation}
\Ham \psi(\vec{r})= \frac{-\hbar^2}{2} \nabla \cdot \left(
\frac{1}{m(\vec{r})} \nabla \psi(\vec{r})\right) +V(\vec{r})
\psi(\vec{r}) . \label{E:Hamiltonian_superlattice}
\end{equation}
Similar to Ref. \cite{silveirinha_giant_2014}, we consider that the
host material is Hg$_{0.75}$Cd$_{0.25}$Te whereas the material of
the inclusions is HgTe. These materials are nearly lattice matched.
Note that unlike the 2DEG studied in the first part of the article,
the unit cell of the HgCdTe superlattice contains only one element.

Following Refs. \cite{jelinek_metamaterial-inspired_2011,
silveirinha_metamaterial-inspired_2012, silveirinha_giant_2014} (see
also Ref. \cite{bastard_wave_1988}) for narrow gap binary compounds
of the groups II-VI the potential $V$ for each bulk material can be
identified with the conduction band energy level ${V}\left( E
\right) = {E_c}$, with ${E_c} = {E_{{\Gamma _6}}}$ the conduction
band edge energy. On the other hand, the dispersive effective mass
may be assumed to be of the form $m(E) = \frac{1}{{2v_P^2}}\left( {E
- {E_v}} \right)$, with ${E_v} = {E_{{\Gamma _8}}}$ the energy level
associated with the edge of the light-hole band and $v_P$ the Kane's
velocity.

For simplicity, here we assume that the elements of our
2D-superlattice can be described by the same parameters as the
corresponding bulk materials. Hence, for an
Hg$_{0.75}$Cd$_{0.25}$Te-HgTe superlattice $V$ in Eq.
\eqref{E:Hamiltonian_superlattice} is such that:
\begin{equation}
 V_h=E_{v,h}+E_g(x=0.25), \qquad  V_i=E_{v,i}+E_g(x=0).
\end{equation}
In the above, $E_g=E_g(x)$ stands for the  band gap energy of the
ternary compound Hg$_{x}$Cd$_{1-x}$Te, which is calculated with
Hansen's formula at zero temperature
\cite{rogalski_hgcdte_2005,hansen_energy_1982}, where $x$ represents
the mole fraction. Notably, the electronic band structure of HgTe is
inverted, so that the conduction ($\Gamma_6$) band (with an $S$-type
symmetry) lies below the valence ($\Gamma_8$) band (with a $P$-type
symmetry), and the band gap energy is negative \cite{Lawaetz,
silveirinha_transformation_electronics}. The valence band offset for
the considered pair of materials can be estimated equal to \cite{
silveirinha_metamaterial-inspired_2012, silveirinha_giant_2014}:
\begin{equation}
E_{v,h}=E_{v,i}-0.0875 \electronvolt.
\end{equation}
The dispersive masses of the relevant semiconductors are
\begin{equation}
 m_h=\frac{1}{2v_P^2}(E-E_{v,h}), \qquad  m_i=\frac{1}{2v_P^2}(E-E_{v,i}).
 \label{E:dispersive_masses}
\end{equation}
where the Kane velocity is supposed to be the same in the two media
$v_P=1.06\cdot 10^6 \meter\cdot\second^{-1}\,\,$
\cite{rogalski_hgcdte_2005}.

\subsection{Effective Hamiltonian and stationary states} \label{sec:effective_parameters_superlattice}

The scalar effective Hamiltonian of the superlattice is computed in
the same way as in section \ref{sec:scalar_Hamiltonian}. Now, BZ
should be taken as the first Brillouin zone because for this
superlattice the Dirac cone emerges at the
$\Gamma$-point\cite{silveirinha_giant_2014}. The details of the
numerical implementation of the FDFD method are described in
Appendix \ref{sec:FDFD_superlattice}.

Interestingly, different from the example of Sect.
\ref{sec:honeycomb_lattice}, our numerical calculations show that
the effective Hamiltonian $H_\text{ef}$ is a smooth function of
$\bf{k}$ at the origin. Hence, it is possible to expand
$H_\text{ef}$ in a Taylor series in $\bf{k}$ as follows:
\begin{equation}
 H_\text{ef}(\k,E) \simeq V_\text{ef}(E) +  \frac{\hbar^2}{2}\k \cdot \db{m}_\text{ef}^{-1}(E) \cdot
 \k, \label{E:Hef_hexagonal}
\end{equation}
where $V_\text{ef}(E)=H_\text{ef}(\k=0,E)$ and the inverse effective
mass tensor is
\begin{equation}
\label{E:Effective_mass_tensor}
\db{m}_\text{ef}^{-1}(E)=\frac{1}{\hbar^2}
 \begin{bmatrix}
 \left. \frac{\partial^2 H_\text{ef}(\k,E)}{\partial k_x^2}\right|_{\vec{k}=0} & \left. \frac{\partial^2 H_\text{ef}(\k,E)}{\partial k_x \partial k_y}\right|_{\vec{k}=0}\\
\\
\left. \frac{\partial^2 H_\text{ef}(\k,E)}{\partial k_x \partial
k_y}\right|_{\vec{k}=0} & \left.\frac{\partial^2
H_\text{ef}(\k,E)}{\partial k_y^2}\right|_{\vec{k}=0}
 \end{bmatrix},
\end{equation}
with $k_x$ and $k_y$ the components of wavevector with respect to
the $x$ and $y$ directions respectively. Note that $H_\text{ef}$ is
an even function of $\k$. In particular, within the validity of Eq.
\eqref{E:Hef_hexagonal} the energy dependent effective Hamiltonian
can be written in the space domain in the form:
\begin{equation}
{\hat H_\text{ef}} =  - \frac{{{\hbar ^2}}}{2}\nabla  \cdot
\db{m}_\text{ef}^{-1}(E) \cdot
 \nabla  + {V_\text{ef}}.
\end{equation}

Based on an analogy with electromagnetic metamaterials, it was found
in Ref. \cite{silveirinha_giant_2014} that the effective mass tensor
and the effective potential of the superlattice may be approximated
by:
\begin{equation}
 m_\text{ef}(E)=m_h\frac{(1-f_V)m_h+(1+f_V)m_i}{(1+f_V)m_h+(1-f_V)m_i},
 \label{E:mef}
\end{equation}
\begin{equation}
 V_\text{ef}=V_h(1-f_V)+V_if_V, \label{E:Vef}
\end{equation}
where $f_V$ represents the volume fraction of the HgTe inclusions.
In the next section, we will compare these analytical formulas with
the results obtained with the numerically calculated
$H_\text{ef}({\bf{k}},E)$.

The dispersion of the electronic states of the superlattice can be
found by solving the secular equation \eqref{E:energy_states}. In
terms of the effective mass tensor and of the potential, it reduces
to
\begin{equation}
  \frac{\hbar^2}{2}\k \cdot \db{m}_\text{ef}^{-1}(E) \cdot \k=E-V_\text{ef}(E).
  \label{E:band_diagram_equation_superlattice_2}
\end{equation}
\subsection{Numerical results}

Using the formalism described in the previous subsection, we
computed the effective parameters and the energy dispersion diagrams
for different Hg$_{0.75}$Cd$_{0.25}$Te-HgTe superlattices with a
lattice constant $a=12 a_s$, where $a_s=0.65 nm$ is the atomic
lattice constant of the semiconductors.
In our previous work, it was predicted that for a critical volume
fraction of the inclusions
\begin{equation}
f_{V_0}  = \frac{{E_{v,h}  + E_{v,i}  - 2V_h }}{{E_{v,h}  - E_{v,i}
- 2\left( {V_h  - V_i } \right)}}
\end{equation}
the superlattice is characterized by a zero band gap at the energy
level $E=V_\text{ef}$, where $V_\text{ef}$ is given by Eq.
\eqref{E:Vef}. For the considered superlattice, $f_{V_0}  =
{\rm{0}}{\rm{.247}}$.

\begin{figure}[!h]
    \centering
 \includegraphics[width=\linewidth]{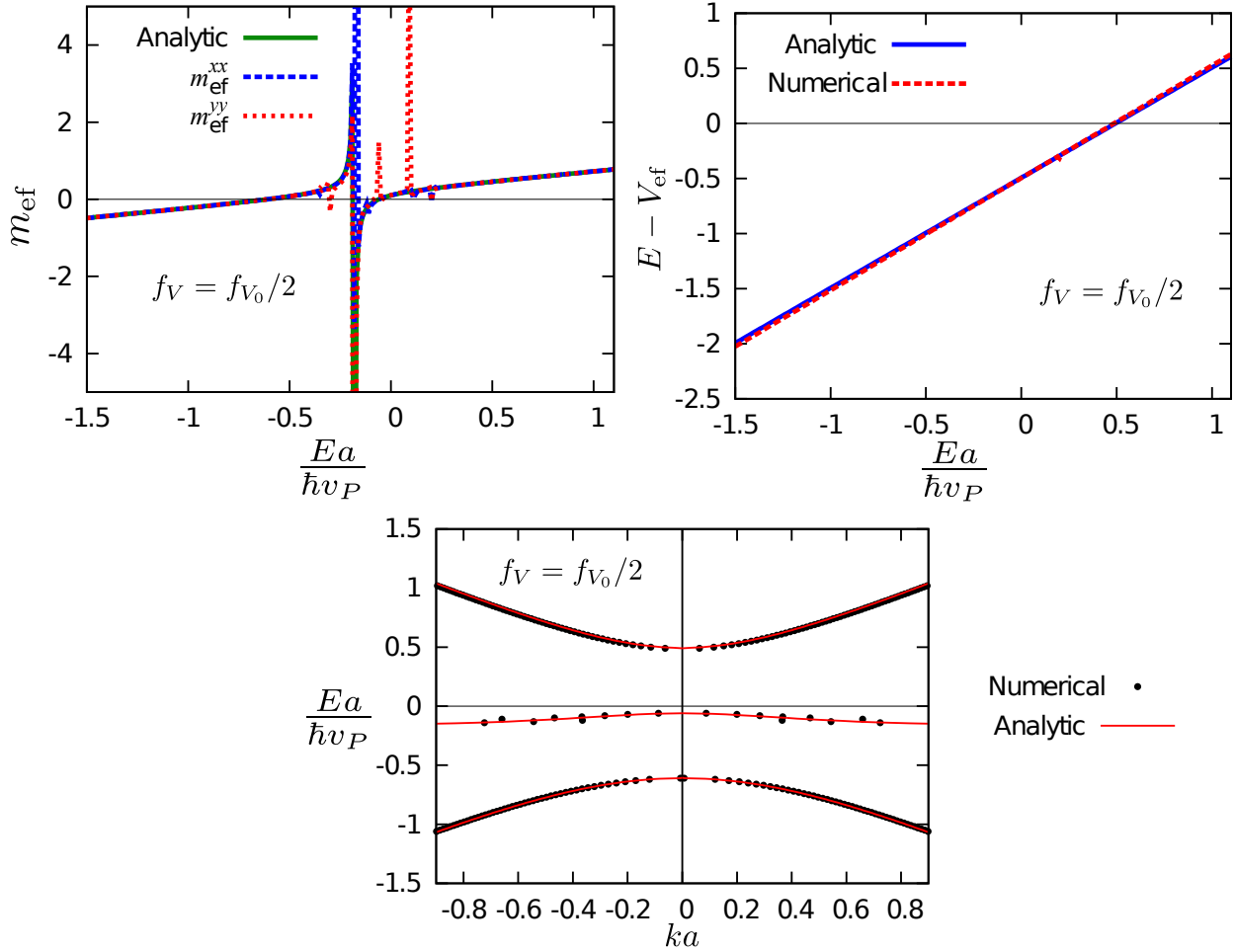}
          \caption{Effective parameters and electronic band structure near the $\Gamma$-point for an HgCdTe superlattice with $f_{V}=f_{V_0}/2$.
    The solid lines represent the analytical results, whereas the discrete symbols/dashed lines are obtained from the numerically calculated effective Hamiltonian.}
\label{fig:band_diagram_superlattice_f_0.1235}
\end{figure}

\begin{figure}[!h]
    \centering
 \includegraphics[width=\linewidth]{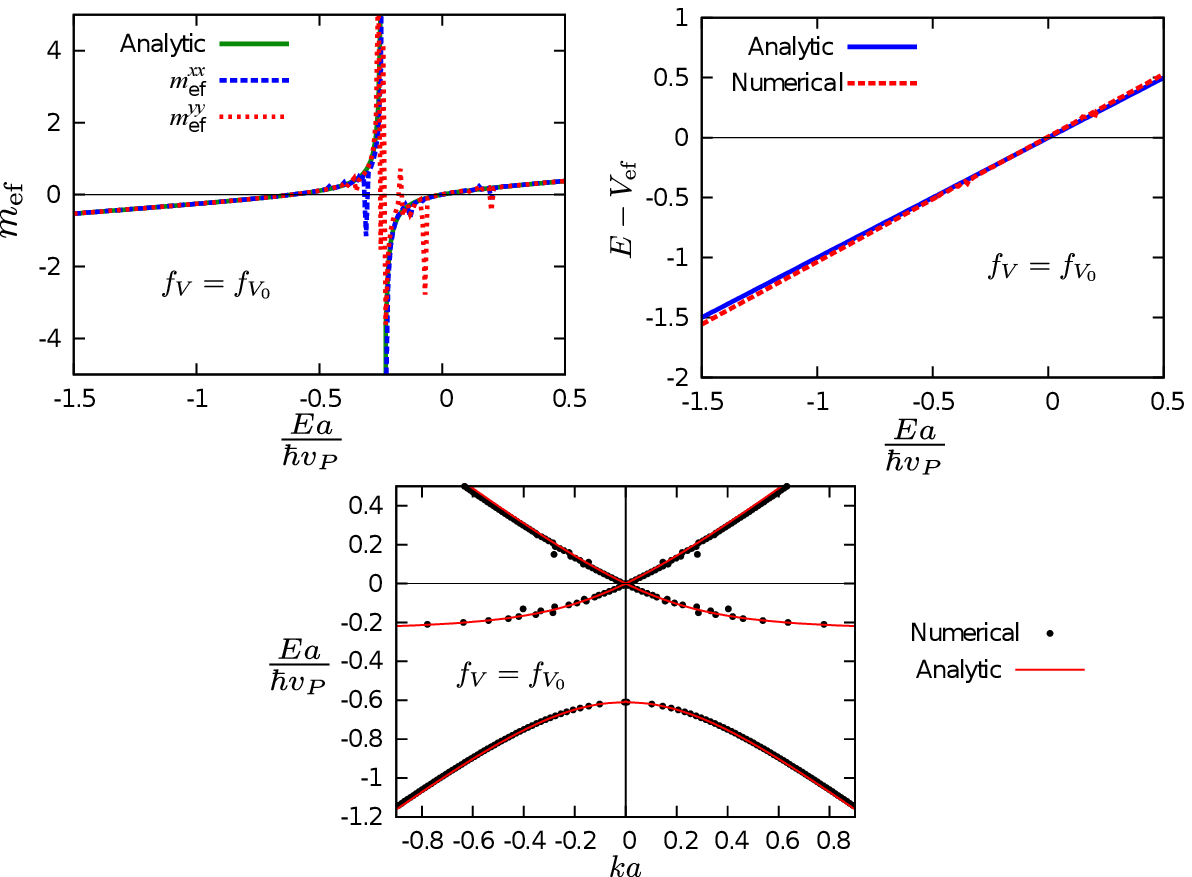}
          \caption{Similar to Fig. \ref{fig:band_diagram_superlattice_f_0.1235} but for $f_{V}=f_{V_0}$.}
\label{fig:band_diagram_superlattice_f_0.247}
\end{figure}

\begin{figure}[!h]
    \centering
 \includegraphics[width=\linewidth]{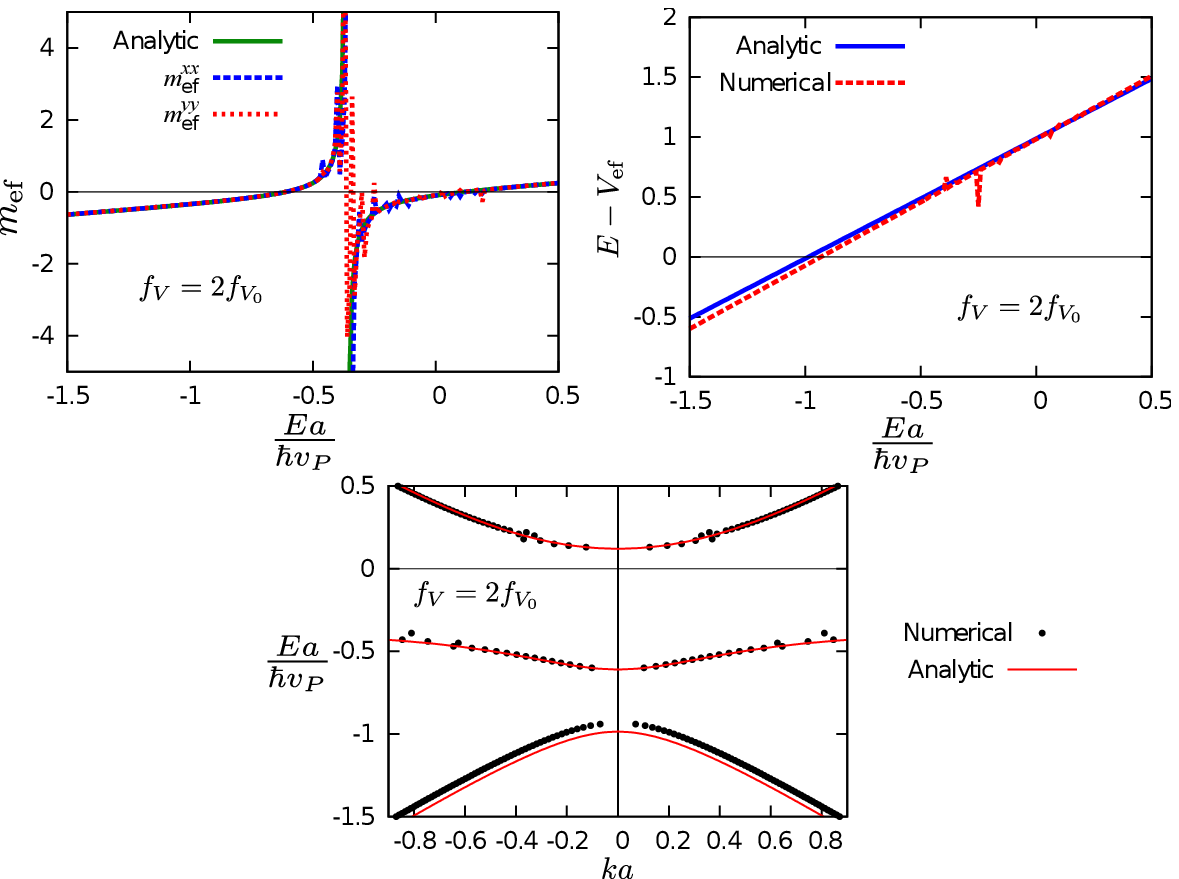}
          \caption{Similar to Fig. \ref{fig:band_diagram_superlattice_f_0.1235} but for $f_{V}=2 f_{V_0}$.}
\label{fig:band_diagram_superlattice_f_0.494}
\end{figure}

Figures \ref{fig:band_diagram_superlattice_f_0.1235},
\ref{fig:band_diagram_superlattice_f_0.247} and
\ref{fig:band_diagram_superlattice_f_0.494} represent the
numerically calculated effective parameters $E-V_\text{ef}$ and
$\db{m}_\text{ef}$ and the energy dispersion for the volume
fractions $f_{V_0}/2$, $f_{V_0}$ and $2f_{V_0}$, respectively. The
out-of-diagonal components of the effective mass tensor are zero,
and hence only the diagonal components are represented in the
Figures. The effective medium results correspond to the discrete
symbols/dashed lines and are superimposed on the results (solid
lines) predicted by the analytical formulas
\eqref{E:Vef}-\eqref{E:mef}. In the simulations, we fixed the energy
scale so that when $f_V = f_{V_0}$ the tip of the Dirac cone is
associated with the energy level $E=0$. This corresponds to choosing
$E_{v,i}$ such that $(E_{v,h}+E_{v,i})V_i-2E_{v,i}V_h=0$
\cite{silveirinha_giant_2014}.
As seen, there is an excellent agreement between the analytic and
effective medium results. Moreover, in Ref.
\cite{silveirinha_giant_2014} it was shown that the analytical
formulas compare very well with exact electronic band structure
calculations based on the plane-wave-method. This demonstrates that
the single component effective Hamiltonian describes correctly the
propagation of electron waves in the HgCdTe superlattice, and thus
that the electrons do not have a pseudospin degree of freedom as in
the modulated 2DEG studied in the first part of the article. A
comparison between Kane-like electrons in semicondutor
heterostructures and Dirac-like electrons in graphene was also
reported in Ref. \cite{Dragoman}.

It is important to mention that the effective medium parameters have
several extra resonances, which that are not predicted by the
analytical formalism. These resonances are associated with
hybridized heavy-hole states, and give rise to extra nearly flat
bands in the electronic band diagrams (not shown). As already
discussed in Ref. \cite{silveirinha_giant_2014}, this property is
the semiconductor counterpart of ``plasmons`` in metallic photonic
crystals \cite{mcgurn_photonic_1993, Costa_ZIM}. The stationary
states of the superlattice occur for the energy levels for which the
effective parameters $E-V_\text{ef}$ and $\db{m}_\text{ef}$ have the
same sign, in agreement with
\eqref{E:band_diagram_equation_superlattice_2}. The two effective
parameters  play a role similar to the permittivity and permeability
in electromagnetics
\cite{jelinek_metamaterial-inspired_2011,silveirinha_metamaterial-inspired_2012}.

As found in Ref. \cite{silveirinha_giant_2014}, the electronic
states of the superlattice associated with the energy
$E=V_\text{ef}$ are electron-like ($\Gamma_6$ band), whereas the
states associated with the energy for which $m_\text{ef}=0$ are
light-hole-like ($\Gamma_8$ band). Thus, the effective medium
results of Figs \ref{fig:band_diagram_superlattice_f_0.1235}-
\ref{fig:band_diagram_superlattice_f_0.494} confirm that for small
values of $f_V$ ($f_V< f_{V_0}$) the superlattice has a normal band
structure similar to the host material, whereas for large values of
$f_V$ ($f_V > f_{V_0}$) the band structure is inverted similar to
HgTe inclusions. The critical volume $f_V=f_{V_0}$ marks the
topological transition from a normal to an inverted band structure,
and is associated with a single valley Dirac cone at the $\Gamma$
point. Because the electrons do not have a pseudospin their time
evolution is not described by a massless Dirac equation. Indeed, in
the present case the linear energy dispersion is not a consequence
of a symmetry of the system, but rather due to the topological band
structure transition. Related band structure transitions have been
reported in HgCdTe quantum wells \cite{Bernevig, Konig}, and mark
the point beyond which the transport associated with edge states
becomes possible.


\section{Conclusion} \label{sec:conclusion}

Using a first principles effective medium approach, it was
demonstrated that the electronic band structure of electron waves in
a 2DEG modulated by an electrostatic potential with honeycomb
symmetry is characterized by the massless 2D Dirac equation near the
corners of the Brillouin zone, exactly as in graphene. Moreover, it
was theoretically shown that the same formalism may also describe
the time evolution of initial ``macroscopic'' electronic states, and
the precise link between the microscopic and effective medium
frameworks was derived. In particular, our theory highlights the
connection between the components of the pseudospinor and the values
of the microscopic wave function in the two sublattices of the 2DEG.
In addition, we characterized HgCdTe superlattices, and demonstrated
that in this second platform the electrons can also have zero
effective mass and linearly dispersing bands. However, different
from the 2DEG artificial graphene, in this second system the
electrons are achiral fermions and a pseudospinor description is not
required. Moreover, the Dirac cone emerges at the $\Gamma$-point and
results from a topological band structure transition, rather than
from the structural symmetry. Finally, we note that the ideas
introduced in this article can be readily extended to photonic
systems, and may enable an effective medium description of
electromagnetic waves in ``photonic graphene''
\cite{segev_graphene}.

\appendix

\section{Calculation of the microscopic wavefunction with the FDFD method}
In order to determine the solution $\psi(\vec{r},\omega)$ of
equation \eqref{E:schrodinger_laplace}, we use the well-known FDFD
method based on the Yee's mesh \cite{yee_numerical_1966}. This
frequency domain method is very well suited to model finite-sized
structures with complex geometries. In this approach the unit cell
is divided into many rectangular subcells and the differential
operators are replaced by finite difference operators on each node
of the mesh.

\subsection{Honeycomb lattice} \label{sec:FDFD_honeycomb_lattice}
Here, we describe the implementation of the FDFD method for the
honeycomb lattice studied in section \ref{sec:honeycomb_lattice}.
 \subsubsection{The unit cell}
The primitive cell of the honeycomb lattice is not rectangular (see
the region delimited by the dotted lines in Fig.
\ref{fig:honeycomb_lattice}), and consequently this cell is not
adequate for the FDFD discretization of Eq.
\eqref{E:schrodinger_laplace}. Hence, to simplify the formulation of
the FDFD problem, we use a rectangular supercell containing four
elements and generated by the vectors  $\vec{a}_1$ and $2
\vec{a}_2-\vec{a}_1$. This supercell is represented by the coloured
area in Fig. \ref{fig:honeycomb_lattice} (left). The coordinates of
the primitive vectors in the Cartesian coordinate system are:
\begin{gather}
\vec{a}_1=a \sqrt{3} \hat{\vec{x}},\\
\vec{a}_2=\frac{\sqrt{3} }{2}a \hat{\vec{x}} + \frac{3}{2}a \hat{\vec{y}},
\end{gather}
where $a$ is the nearest neighbor distance .\\
The coordinates of the reciprocal lattice vectors $\vec{b}_1$ and
$\vec{b}_2$ represented in the Figure \ref{fig:honeycomb_lattice}
(right) are
\begin{gather}
\vec{b}_1= \dfrac{2\pi}{a\sqrt{3}}\left( \hat{\vec{x}} -\frac{1}{\sqrt{3}} \hat{\vec{y}}  \right),\\
\vec{b}_2= \dfrac{4\pi}{3a} \hat{\vec{y}}.
\end{gather}
 \subsubsection{FDFD discretization}

\begin{figure}[!h]
    \centering
 \includegraphics[width=0.35\linewidth]{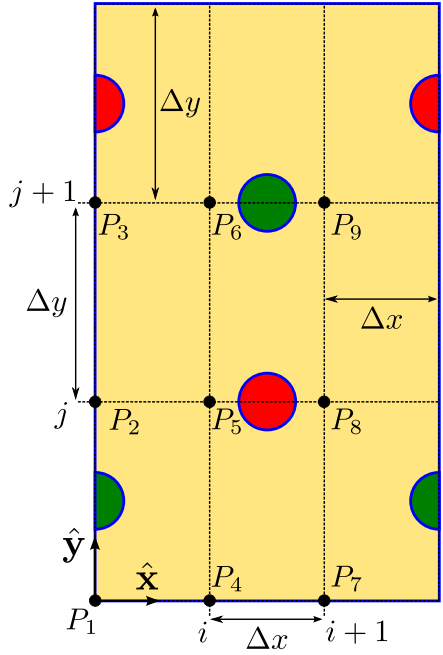}
         \caption{Geometry of the grid mesh for the FDFD discretization of the supercell in the particular case $N_x=N_y=3$.}
\label{fig:unit_cell_FDFD}
\end{figure}

For an initial macroscopic state $-i\hbar
\psi_{t=0}(\vec{r})=\e{i\k\cdot\vec{r}}$ (playing the role of a
source), the problem to be solved [Eq.
\eqref{E:schrodinger_laplace}] reduces to
\begin{equation}
 \left[\frac{-\hbar^2}{2m_b} \left( \diffsec{}{x} + \diffsec{}{y}\right)+V(x,y)-E \right] \psi(\vec{r}) = \e{i\k\cdot\vec{r}}.
\label{E:schrodinger_laplace_2}
\end{equation}
As illustrated in Fig. \ref{fig:unit_cell_FDFD}, the supercell is
discretized using a uniform grid, with $N_x$ and $N_y$ nodes along
the $x$ and $y$ directions, respectively. The grid spacing along the
$x$ and $y$ directions are $\Delta x=\dfrac{a\sqrt{3}}{N_x}$ and
$\Delta y=\dfrac{3a}{N_y}$.
The differential operators are discretized as
\cite{yasumoto_electromagnetic_2010, Costa_homogenization}
\begin{gather}
\diffsec{\psi(i,j)}{x}= \dfrac{\psi(i+1,j)-2\psi(i,j)+\psi(i-1,j)}{(\Delta x)^2}\\
\diffsec{\psi(i,j)}{y}=
\dfrac{\psi(i,j+1)-2\psi(i,j)+\psi(i,j-1)}{(\Delta y)^2}
\end{gather}
Hence, Eq. \eqref{E:schrodinger_laplace_2} reduces to:
\begin{multline}
\frac{-\hbar^2}{2m_b}\left(
\dfrac{\psi(i+1,j)-2\psi(i,j)+\psi(i-1,j)}{(\Delta x)^2} +
\dfrac{\psi(i,j+1)-2\psi(i,j)+\psi(i,j-1)}{(\Delta y)^2}\right) \\ +
V(i,j)\cdot \psi(i,j) -E \cdot \psi(i,j) = \e{i\k\cdot\vec{r}(i,j)},
\label{E:FDFD}
\end{multline}
where $\vec{r}(i,j)$ refers to the vector position at the node
$(i,j)$ of the grid (see Fig. \ref{fig:unit_cell_FDFD}). For the
nodes located at the boundaries of the supercell, one or more
adjacent node(s) may lie outside the grid. However, these nodes can
be brought back inside the supercell using the Bloch-periodic
boundary conditions:
\begin{gather}
\psi(x+ \|\vec{a}_1\| ,y)= \psi(x+a\sqrt{3} ,y) =  \psi(x,y) \e{i k_x a\sqrt{3} }, \\
\psi(x,y+ \| 2 \vec{a}_2-\vec{a}_1 \| )  =\psi(x,y+3a)  =  \psi(x,y)
\e{i 3k_y a}, \label{E:periodic_boundary_conditions}
\end{gather}
where $\| ~\|$ refers to the norm of a vector, and $k_x$, $k_y$ are
the components of the wave vector along the $x$ and $y$ directions.

The system of equations \eqref{E:FDFD} is equivalent to a standard
linear system with  $N_x \times N_y$ equations and $N_x \times N_y$
unknowns [the $\psi(i,j)$]. This system can be solved with standard
numerical methods, and its solution can then be used to compute the
effective Hamiltonian as explained in section
\ref{sec:honeycomb_lattice}.


\subsection{Hexagonal superlattice} \label{sec:FDFD_superlattice}

The FDFD method implementation for the hexagonal superlattice
studied in section \ref{sec:triangular_superlattice} is very similar
to what was described in the previous subsection. In this case, we
used a supercell generated by the primitive vectors $2\vec{a}_1
-\vec{a}_2$ and $\vec{a}_2$ (see the coloured area in Fig.
\ref{fig:superlattice}), with $\vec{a}_1=\frac{\sqrt{3} }{2}a
\hat{\vec{x}} + \frac{1}{2}a \hat{\vec{y}}$ and $\vec{a}_2=a
\hat{\vec{y}}$ and $a$ the lattice constant.
For an initial macroscopic state $-i\hbar
\psi_{t=0}(\vec{r})=\e{i\k\cdot\vec{r}}$ the equation to be solved
is
\begin{equation}
 \frac{-\hbar^2}{2} \left[ \frac{\partial}{\partial x} \left( \frac{1}{m(\vec{r})} \frac{\partial\psi(\vec{r})}{\partial x}  \right) + \frac{\partial}{\partial y}\left( \frac{1}{m(\vec{r})}  \frac{\partial\psi(\vec{r})}{\partial y} \right) \right] +V(\vec{r}) \psi(\vec{r}) -E  \psi(\vec{r}) = \e{i\k\cdot\vec{r}}.
 \label{E:schrodinger_laplace_superlattice2}
\end{equation}
The differential operators are discretized by using staggered subgrids for $\psi$, $\partial_x\psi$ and $\partial_y\psi$ similar to Yee's approach \cite{yee_numerical_1966}. This gives:
\begin{gather}
 \frac{\partial}{\partial x} \left( \frac{1}{m(i,j)} \frac{\partial\psi(i,j)}{\partial x}  \right)  = \frac{1}{(\Delta x)^2} \left( \dfrac{\psi(i+1,j)-\psi(i,j)}{m(i+\frac{1}{2},j)} -\dfrac{\psi(i,j)-\psi(i-1,j)}{m(i-\frac{1}{2},j)} \right)  \\
\frac{\partial}{\partial y}\left( \frac{1}{m(i,j)}  \frac{\partial\psi(i,j)}{\partial y} \right) = \frac{1}{(\Delta y)^2} \left( \dfrac{\psi(i,j+1)-\psi(i,j)}{m(i,j+\frac{1}{2})} - \dfrac{\psi(i,j)-\psi(i,j-1)}{m(i,j-\frac{1}{2})} \right)
\end{gather}
where the grid spacing along the $x$ and $y$ directions are $\Delta x=\dfrac{a\sqrt{3}}{N_x}$ and $\Delta y=\dfrac{a}{N_y}$.
In this manner, the problem is reduced to a $N_x \times N_y$ linear system, analogous to the previous subsection.
%
%


\section{Probability density function in the effective medium approach} \label{sec:Probability_density_pseudospinor}

In this appendix, we demonstrate the relation
\eqref{E:density_probability_average_psi} that links the microscopic
and macroscopic probability densities. This result extends those of
Refs. \cite{fernandes_wormhole, silveirinha_poynting_2009,
Costa_poynting}.

For convenience, we denote $\Phi_e= \left\{ \Phi \right\}_\text{av}$ and introduce the inner product
\begin{equation}
\left\langle \Phi_1 \left| \Phi_2 \right. \right\rangle =
\frac{1}{V_c}\int_\Omega \Phi_1^\ast \cdot \Phi_2 \, \Dr[N]{\vec{r}}
.
\end{equation}
Moreover, it is easy to show that for a Bloch mode $g_\k$ with $\k$
in the Brillouin zone we have the following property
\begin{equation}
\left\langle g_\k \left| g_\k \right. \right\rangle = \left\{
g_\k^\ast \cdot g_\k\right\}_\text{av}.
\label{E:properties_average_inner_product}
\end{equation}
The starting point is to consider a family of solutions $\Phi=\Phi(\vec{r},E)$ of Eq. \eqref{E:Laplace_generalized} parameterized by the energy $E$, and for initial conditions such that $f_1=f_2=f_0$. Thus, $\Phi=\Phi(\vec{r},E)$ satisfies:
\begin{equation}
\left( \Ham_g-E \right)  \cdot \Phi=
   \begin{pmatrix} \chi_1 & 0 \\0 & \chi_2   \end{pmatrix} \begin{pmatrix} 1 \\ 1  \end{pmatrix}
 f_0\e{i\k\cdot\vec{r}}.
 \label{E:schrodinger_pseudospinor_source}
\end{equation}
From Sect. \ref{Sec:subsect_pseudospinor} and from the definition of the effective Hamiltonian, it is evident that:
\begin{equation}
\left( H_{g,\text{ef}}(\k,E)-E \right) \cdot \Phi_e = \frac{1}{2}
 \begin{pmatrix} 1 \\ 1  \end{pmatrix} f_0  \e{i\k\cdot\vec{r}},
 \label{E:schrodinger_pseudospinor_average_source}
\end{equation}
Hence, by combining the two equations it is possible to write:
\begin{equation}
\left( \Ham_g-E \right)  \cdot \Phi= 2
   \begin{pmatrix} \chi_1 & 0 \\0 & \chi_2   \end{pmatrix} \cdot \left( H_{g,\text{ef}}(\k,E)-E \right) \cdot \Phi_e .
   \label{E:equality_normal_average_schrodinger_pseudospinor}
\end{equation}
Differentiating both sides with respect to $E$ and taking the inner
product of the resulting equation with $\Phi$, we get
\begin{multline}
\left\langle \Phi \left| \left( \Ham_g-E \right) \cdot \frac{\partial
\Phi }{\partial E} \right. \right\rangle - \left\langle \Phi | \Phi
\right\rangle \\ = 2 \left\langle \Phi \left|
   \begin{pmatrix} \chi_1 & 0 \\0 & \chi_2   \end{pmatrix} \cdot  [H_{g,\text{ef}}(\k,E)-E] \cdot \frac{\partial \Phi_e }{\partial E} \right. \right\rangle - 2 \left\langle \Phi  \left|
   \begin{pmatrix} \chi_1 & 0 \\0 & \chi_2   \end{pmatrix} \cdot  \left( 1- \frac{\partial H_{g,\text{ef}}(\k,E)}{\partial E} \right) \cdot \Phi_e  \right. \right\rangle
\label{E:scalar_product_derivative_schrodinger_pseudospinor}
\end{multline}
Using the fact that $\begin{pmatrix} \chi_1 & 0 \\0 & \chi_2
\end{pmatrix} \cdot \Phi= \Phi$ and Eq.
\eqref{E:psi_average_pseudospinor}, it is found that
\begin{align}
\left\langle \Phi  \left|
   \begin{pmatrix} \chi_1 & 0 \\0 & \chi_2   \end{pmatrix} \cdot \left( 1- \frac{\partial H_{g,\text{ef}}(\k,E)}{\partial E} \right) \cdot \Phi_e  \right. \right\rangle
    &=\left( \frac{1}{V_c}\int_\Omega \Phi^\ast \e{i\k\cdot\vec{r}} \, \Dr[N]{\vec{r}} \right) \cdot \left( 1- \frac{\partial H_{g,\text{ef}}(\k,E)}{\partial E} \right) \cdot \Phi_\text{av}  \nonumber \\
  &= \Phi_e^\ast \cdot  \left( 1- \frac{\partial H_{g,\text{ef}}(\k,E)}{\partial E} \right) \cdot \Phi_e,
\end{align}
Based on similar arguments, it is possible to verify that
\begin{equation}
\left\langle \Phi \left|
   \begin{pmatrix} \chi_1 & 0 \\0 & \chi_2   \end{pmatrix} \cdot  [H_{g,\text{ef}}(\k,E)-E] \cdot \frac{\partial \Phi_e }{\partial E} \right. \right\rangle = \Phi_e^\ast \cdot [H_{g,\text{ef}}(\k,E)-E] \cdot \frac{\partial \Phi_e }{\partial E}.
\end{equation}
On the other hand, because $\Ham_g$ is Hermitian (thus $H_{g,\text{ef}}$ is also an Hermitian matrix) and from Eq.
\eqref{E:equality_normal_average_schrodinger_pseudospinor}, it follows that:
\begin{align}
\left\langle \Phi \left| \left( \Ham_g-E \right) \frac{\partial  \Phi }{\partial E} \right. \right\rangle
&= \left\langle 2
   \begin{pmatrix} \chi_1 & 0 \\0 & \chi_2   \end{pmatrix} \cdot \left( H_{g,\text{ef}}(\k,E)-E \right) \cdot \Phi_e  \left|  \frac{\partial  \Phi }{\partial E} \right. \right\rangle \nonumber \\
&= 2 \left\langle  \Phi_e  \left|
     \left( H_{g,\text{ef}}(\k,E)-E \right) \cdot \begin{pmatrix} \chi_1 & 0 \\0 & \chi_2   \end{pmatrix} \cdot \frac{\partial  \Phi }{\partial E} \right. \right\rangle \nonumber \\
&=2 \Phi_e^\ast \cdot [H_{g,\text{ef}}(\k,E)-E] \cdot \frac{\partial
\Phi_e }{\partial E}
\end{align}
Then by substitution of the previous results into
\eqref{E:scalar_product_derivative_schrodinger_pseudospinor}, we
conclude that
\begin{equation}
 \left\langle \Phi | \Phi  \right\rangle  = 2  \Phi_e^\ast \cdot  \left( 1- \frac{\partial H_{g,\text{ef}}(\k,E)}{\partial E} \right) \cdot \Phi_e.
\end{equation}
This result and Eq. \eqref{E:properties_average_inner_product} yield
\eqref{E:density_probability_average_psi}, as we wanted to prove.
Note that because $f_0$ in Eq.
\eqref{E:schrodinger_pseudospinor_average_source} is an arbitrary
function of the energy, the derived result applies to any solution
of \eqref{E:schrodinger_pseudospinor_average_source}, and in
particular to the electronic stationary states ($f_0(E)=0$).

\section{The matrices $\tilde{\vec{A}}_1$ and
$\tilde{\vec{A}}_2$} \label{sec:ApC}

To illustrate the accuracy of the identities in Eq. \eqref{E:Atil},
we consider the particular case wherein the structural parameters of
the 2DEG satisfy $V_0=-0.8~\milli\electronvolt$ and $R/a=0.35$, as
in Fig. \ref{fig:comparison_single_pseudospinor}. The numerically
computed matrices are such that (showing 3 significant figures):
\begin{equation}
\frac{1}{\hbar v_F \cos \phi} \Re(\tilde{\vec{A}}_1)=\begin{pmatrix}
-5.39\cdot 10^{-3} & 0.999\\ 0.999 & 8.97\cdot 10^{-3}
\end{pmatrix},
\end{equation}
\begin{equation}
\frac{1}{\hbar v_F \sin \phi} \Re(\tilde{\vec{A}}_2)=
\begin{pmatrix} 6.62\cdot 10^{-3} & 1.00 \\ 1.00 & -6.03\cdot
10^{-3} \end{pmatrix},
\end{equation}
\begin{equation}
\frac{-1}{\hbar v_F \sin \phi} \Im(\tilde{\vec{A}}_1)=
\begin{pmatrix} -7.84\cdot 10^{-11} & -0.999  \\ 0.999  &
-8.35\cdot10^{-10} \end{pmatrix},
\end{equation}
\begin{equation}
\frac{1}{\hbar v_F \cos \phi} \Im(\tilde{\vec{A}}_2)=
\begin{pmatrix} -2.26\cdot10^{-11}  & -1.00  \\ 1.00 &
2.05\cdot10^{-9} \end{pmatrix},
\end{equation}
with $\phi = 60\degree$ and $v_F =9.03\cdot10^{3}
~\meter\second^{-1} $. These results demonstrate that Eq.
\eqref{E:Atil} is very accurate.
%
%


\bibliographystyle{ieeetr}

\bibliography{Biblio_v5}

\end{document}